\documentclass[%
 aip,
 amsmath,amssymb,
 reprint,%
]{revtex4-2}

\usepackage{graphicx}
\usepackage{dcolumn}
\usepackage{bm}

\usepackage[utf8]{inputenc}
\usepackage[T1]{fontenc}
\usepackage{mathptmx}
\usepackage{etoolbox}

\makeatletter
\def\@email#1#2{%
 \endgroup
 \patchcmd{\titleblock@produce}
  {\frontmatter@RRAPformat}
  {\frontmatter@RRAPformat{\produce@RRAP{*#1\href{mailto:#2}{#2}}}\frontmatter@RRAPformat}
  {}{}
}%

\usepackage{lipsum}
\usepackage{color}
\usepackage{amsmath}
\usepackage{color}
\usepackage{siunitx}
\usepackage{float}
\usepackage{soul}
\usepackage{pifont,wasysym}
\usepackage{color,amsmath}
\usepackage{mathrsfs} 
\usepackage{amsmath}

\def\boldsymbol{\bm}

\usepackage[unicode=true,
  linktocpage,
  linkbordercolor={0.5 0.5 1},
  citebordercolor={0.5 1 0.5},
  linkcolor=blue, citecolor = blue, colorlinks=true]{hyperref}

\definecolor{dgreen}{rgb}{0.0, 0.4, 0.0}

\usepackage{lipsum} 
\makeatother
\begin{document}

\title{The effect of axisymmetric confinement on propulsion of a three-sphere microswimmer}

\author{Ali G\"urb\"uz}
\affiliation{Department of Mechanical Engineering,
Santa Clara University, Santa Clara, California, USA}

\author{Andrew Lemus}
\affiliation{Department of Mechanical Engineering,
Santa Clara University, Santa Clara, California, USA}

\author{Ebru Demir}
\affiliation{Department of Mechanical Engineering and Mechanics, Lehigh University, Bethlehem, Pennsylvania, USA}

\author{On Shun Pak}\thanks{Electronic mail: opak@scu.edu}
\affiliation{Department of Mechanical Engineering,
Santa Clara University, Santa Clara, California, USA}

\author{Abdallah Daddi-Moussa-Ider}\thanks{Electronic mail: abdallah.daddi-moussa-ider@ds.mpg.de }
\affiliation{Max Planck Institute for Dynamics and Self-Organization, Göttingen, Germany} 

\date{\today}

\begin{abstract}
Swimming at the microscale has recently garnered substantial attention due to the fundamental biological significance of swimming microorganisms and the wide range of biomedical applications for artificial microswimmers. These microswimmers invariably find themselves surrounded by different confining boundaries, which can impact their locomotion in significant and diverse ways. In this work, we employ a widely used three-sphere swimmer model to investigate the effect of confinement on swimming at low Reynolds numbers. We conduct theoretical analysis via the point-particle approximation and numerical simulations based on the finite element method to examine the motion of the swimmer along the centerline in a capillary tube. The axisymmetric configuration reduces the motion to one-dimensional movement, which allows us to quantify how the degree of confinement affects the propulsion speed in a simple manner. Our results show that the confinement does not significantly affect the propulsion speed until the ratio of the radius of the tube to the radius of the sphere is in the range of $\mathcal{O}(1)-\mathcal{O}(10)$, where the swimmer undergoes substantial reduction in its propulsion speed as the radius of the tube decreases. We provide some physical insights into how reduced hydrodynamic interactions between moving spheres under confinement may hinder the propulsion of the three-sphere swimmer. We also remark that the reduced propulsion performance stands in stark contrast to the enhanced helical propulsion observed in a capillary tube, highlighting how the manifestation of confinement effects can vary qualitatively depending on the propulsion mechanisms employed by the swimmers.
\end{abstract}

\maketitle

\section{Introduction} \label{sec:Intro}
The study of locomotion in fluids at the microscopic scale has attracted significant attention in recent decades. This growing interest is not only driven by the motivation to better understand the motility of swimming microorganisms \cite{Brennen1977,Bray2000,EricLauga2020} but also the potential biomedical applications of artificial microswimmers such as targeted drug delivery and minimally invasive microsurgery \cite{Nelson2010,Sitti2015,JinxingLi2017,WenqiHu2018, Palagi2018,Tsang2020}. Locomotion of biological and artificial microswimmers occurs at negligibly small Reynolds numbers (Re), where viscous forces largely dominate inertial forces. In the inertialess regime, the ability to self-propel is severely constrained owing to kinematic reversibility. In particular, Purcell's scallop theorem \cite{Purcell1977} states that in the absence of inertia, deformations exhibiting time-reversal symmetry (e.g., the motion of a single-hinged scallop opening and closing its shell), also known as reciprocal motion, are unable to produce any net self-propulsion. Common macroscopic swimming strategies such as rigid flapping motion hence become largely ineffective at low Re. Microorganisms such as bacteria and spermatozoa have evolved strategies that utilize biological appendages called flagella with the action of molecular motors to 
swim in their microscopic world. Extensive studies in the past decades have elucidated the physical principles underlying their motility \cite{Fauci06,LaugaPowers2009,Yeomans2014,bechinger2016active,Lauga2016,Wan2022}. 

In parallel efforts, researchers have sought simple and effective mechanisms to develop artificial microswimmers \cite{Abbott2009,Ebbens2010,Lauga2011,sharan2023pair}. In his pioneering work, Purcell demonstrated how a three-link swimmer \cite{Purcell1977}, now known as Purcell’s swimmer \cite{becker2003self,tam2007optimal,Avron2008,Qin2023_B,Qin2023}, can generate net translation with kinematically irreversible cyclic motions. This elegant example has inspired subsequent development of mechanisms that can overcome the fundamental challenge of generating self-propulsion in the inertialess regime. In particular, Najafi and Golestanian \cite{Najafi2004} developed a swimmer consisting of three spheres connected by two extensible rods, which adjust their lengths in a cyclic manner to ingeniously exploit hydrodynamic interactions between the spheres for self-propulsion. The mechanism has also engendered a variety of variants \cite{Avron2005, Earl2007, Golestanian2009, Alouges2008, Alouges2013B, Montino2015,Wang2018, Wang2019, Nasouri2019,Liu2021,Berdakin2022} and their experimental realizations \cite{Leoni2009, Grosjean2016, Box2017,Silverberg_2020}. For its simplicity, the three-sphere swimmer has gained popularity as a useful model for examining different fundamental aspects of locomotion at low Re, including the effect of complex rheology \cite{Curtis2013,yasuda2023generalized}, optimized locomotion \cite{Wang2019,Nasouri2019}, interactions of swimmers \cite{Pooley2007,Farzin2012,yasuda2023generalized}, and swimming near walls \cite{Zargar2009,Najafi2013,Daddi-Moussa-Ider_2018, DaddiMoussaIder2018_2}. 
The three-sphere model has further been used to investigate the reorientation dynamics of microswimmers with respect to flow gradients (rheotaxis)~\cite{daddi2020tuning}, finding that payloads can be exploited to enhance their motion against flows.
More recently, the model has also been employed to explore the integration with machine learning in realizing smart microswimmers \cite{AlanChengHouTsang2020, Hartl2021,Zou2022,Paz2023,Liu2023}. 

\begin{figure*}[t]
    \centering
    \includegraphics[width=0.95\textwidth]{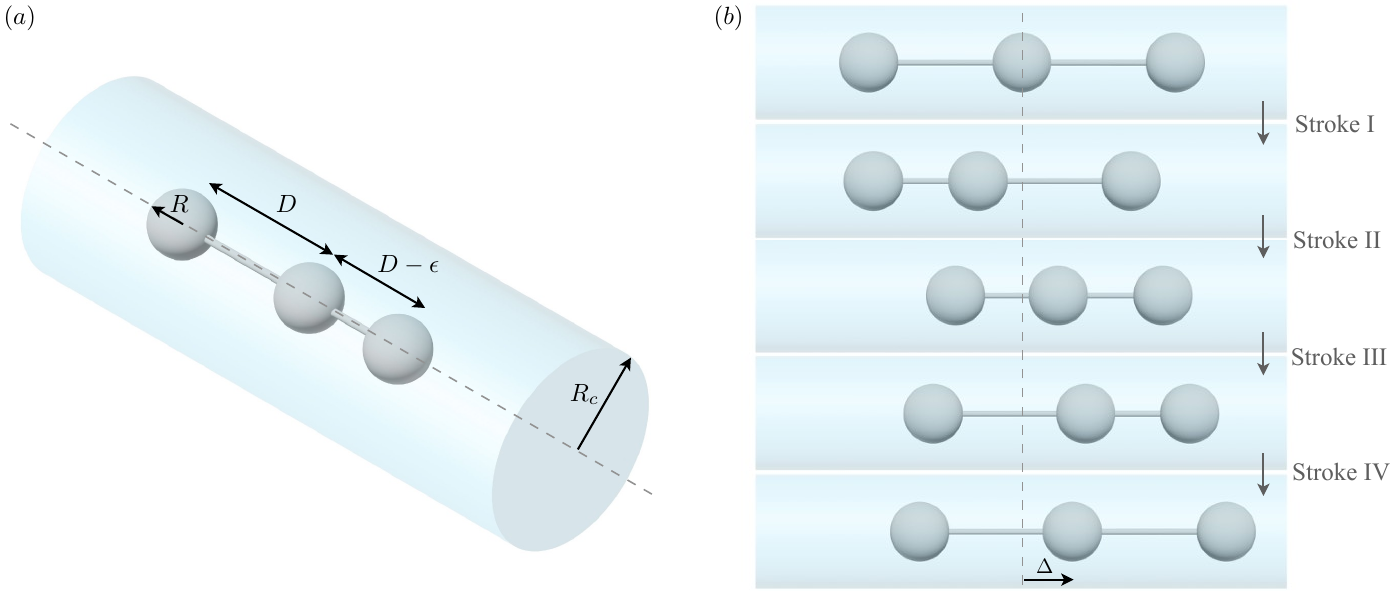}
    \caption{Schematic of the problem setup and notations. (a) A swimmer consisting of three spheres of equal radii $R$ connected by two extensible rods is confined axisymmetrically in a capillary tube of radius $R_c$. The rods have a fully extended length of $D$ and a fully contracted length of $D-\epsilon$, where $\epsilon$ denotes the amount of contraction. (b) The swimmer undergoes a four-stroke cycle designed by Najafi and Golestanian \cite{Najafi2004} to produce a net displacement, $\Delta$.} 
    \label{fig1}
\end{figure*}

Here we utilize the three-sphere swimmer model to probe the effect of confinement on swimming at low Re. Microswimmers invariably find themselves surrounded by different confining boundaries. Extensive studies have demonstrated how swimming near planar boundaries can impact locomotion in significant and diverse manners \cite{Zargar2009,Najafi2013,Daddi-Moussa-Ider_2018, DaddiMoussaIder2018_2, katz_1974,Fauci1995,Ortiz2005,Lauga2006,smith2009,Davide2010,Shum2010,Crowdy2010,Andreas2012,spagnolie_lauga_2012,Li2014,Bayati2019,Farutin2019}. Microorganisms also encounter more complex confinements than planar boundaries, such as spermatozoa swimming through fallopian tubes, parasites \textit{Trypanosomes} in blood vessels, and bacterial motion in microporous soil environments. For swimming inside a capillary tube, previous studies have shown that whether the confinement enhances or hinders propulsion largely depends on the type of swimmers \cite{Felderhof2010,Jana2012,Zhu2013,LedesmaAguilar2013,Liu2014,Caldag2019,ouyang_lin_yu_lin_phan-thien_2022}. For instance, rotating helical flagella, which generate propulsion as a result of drag anisotropy of the slender flagella, display enhanced propulsion speeds inside a capillary tube \cite{Liu2014}. However, squirmers with distribution of tangential surface velocities always have reduced propulsion speeds \cite{Zhu2013}. These results suggest that the difference in the propulsion mechanism of the swimmers play significant roles in how the confinement impacts propulsion. In this work, we consider another physically different propulsion mechanism, namely the three-sphere swimmer model, which relies on hydrodynamic interaction between the spheres for self-propulsion. For simplicity, we focus on the effect of axisymmetric confinement when the swimmer self-propels along the center-line inside a capillary tube. We use both the point-particle approximation and finite element method to quantify how the degree of confinement affects the propulsion speed of this widely used swimmer model. We also provide some physical insights into the underlying mechanism by which confinement influences this specific mode of propulsion.

This paper is organized as follows. We formulate the problem in Sec.~\ref{sec:Formulation} by presenting the swimmer model, the geometrical setup, and the methods of analysis. In Sec.~\ref{sec: Results and discussion}, we first validate our theoretical and numerical results by revisiting the case of an unbounded fluid domain (Sec.~\ref{sec:unbounded}), before discussing new results for confined swimming (Sec.~\ref{sec:Confinement}). We conclude this study in Sec.~\ref{sec:conclusion} with remarks on its limitations and potential directions for future studies.

\section{Problem formulation} \label{sec:Formulation}

\subsection{Swimmer model}\label{sec:SwimmerModel}
We consider the motion of a three-sphere microswimmer confined axisymmetrically in a capillary tube of radius $R_c$. The swimmer was first studied in an unbounded fluid domain by Najafi and Golestanian \cite{Najafi2004}. As illustrated in Fig.~\ref{fig1}(a), the swimmer consists of three spheres of the same radius $R$ connected by two extensible rods of negligible hydrodynamic influences. The fully extended length of each arm is given by $D$ and the fully contracted length of each arm is given by $D-\epsilon$, where $\epsilon$ denotes the amount of contraction or extension in each stroke (referred to as the contraction length hereafter). In the main text, we follow Najafi and Golestanian \cite{Najafi2004} to consider a constant relative speed $W$ in the change of the arm length in the four strokes illustrated in Fig.~\ref{fig1}(b): In stroke I, the swimmer contracts its left arm of an initial length $D$ by an amount $\epsilon$, keeping the length of the right arm at $D$.  In stroke II, the swimmer contracts its right arm by an amount $\epsilon$, keeping the length of the left arm at $D-\epsilon$. In stroke III, the swimmer extends its left arm to reach the fully extended length $D$, with the length of the right arm fixed at $D-\epsilon$. Finally, in stroke IV, the swimmer extends its right arm to return to its original configuration with both arms fully extended with length $D$, completing a full swimming cycle. The net displacement generated by such a cycle is denoted by $\Delta$. In addition to these original strokes considered by Najafi and Golestanian \cite{Najafi2004}, harmonic variations of the length of the two rods have been analyzed in subsequent works \cite{GolestanianAjdari2008}. We conduct the same analyses for the case of harmonic deformations of the rods in the Appendix to assess the generality of our findings.

\subsection{Theoretical analysis: Point-particle approximation}
\label{sec:point-particle}

The motion of an incompressible flow in a Newtonian fluid at low Re is governed by the Stokes equation
\begin{align}
     \mu\nabla^2\textbf{u} &=\nabla p, \label{eqn:Stokes1} \\
     \nabla\cdot\textbf{u} &=0, \label{eqn:Stokes2}
\end{align}
where $\mu$ is the dynamic viscosity, and $\mathbf{u}$ and $p$ are, respectively, the fluid velocity and pressure fields. We denote the velocity of the $i$-th sphere as $\mathbf{V}_i$ and the force and torque acting on them as $\mathbf{F}_i$ and $\mathbf{T}_i$, respectively. No-slip boundary condition are applied on the spheres and the confining tube, \textit{i.e.} $\mathbf{u}_{\text{on the $i$-th sphere}}=\mathbf{V}_i$ and  $\mathbf{u}_{\text{on the confining tube}}=\mathbf{0}$. Without external forces and external torques, the system should be force-free,
\begin{align}
     \sum_{i=1}^{3}{\textbf{F}_i=\textbf{0}}, \label{eqn:Forcefree}
\end{align}
and torque-free,
\begin{align}
     \sum_{i=1}^{3}{\textbf{T}_i=\textbf{0}}. \label{eqn:Torquefree}
\end{align}
As a remark, the torque-free condition is identically satisfied by symmetry  of the problem setup.

We denote by~$\mathbf{r}_1$ the position of the center sphere, which is chosen as a reference for tracking the movement of the swimmer.
We denote by $\mathbf{r}_2$ and~$\mathbf{r}_3$ the positions of front and rear spheres, respectively.
The temporal change in the mutual distances between the spheres is set to perform a non-reversible time sequence.
Under the action of the internally generated forces acting between the spheres along the tube axis ($z$-axis), the lengths of the rod connecting adjacent spheres are set as
\begin{equation}
    \left( \mathbf{r}_2 - \mathbf{r}_1 \right) \cdot \hat{\mathbf{e}}_z = g(t) \, , \qquad
    \left( \mathbf{r}_1 - \mathbf{r}_3 \right) \cdot \hat{\mathbf{e}}_z = h(t) \, ,  
    \label{eq:gh}
\end{equation}
where 
$(g,h) = \left( D,D-Wt \right)$ for $ t \in [0,T/4]$, 
$(g,h) = \left( D+\epsilon-Wt, D-\epsilon \right)$ for $t \in [T/4,T/2]$, 
$(g,h) = \left( D-\epsilon, D-3\epsilon+Wt \right)$ for $t \in [T/2, 3T/4]$, $(g,h) = \left( D-4\epsilon+Wt, D \right)$ for $t \in [3T/4,T]$, and $\hat{\mathbf{e}}_z$ is the unit vector along the $z$-direction.

At low Re, inertial effects are negligible so that the immersed particles take on the velocity of the surrounding fluid instantaneously. 
Accordingly, the translational velocities of the three spheres are related to the internal forces exerted on them linearly via
\begin{equation}
    \mathbf{V}_i = \frac{\mathrm{d} \mathbf{r}_i}{\mathrm{d} t} 
    = \sum_{j=1}^3 \boldsymbol{\mu}_{ij} \cdot \mathbf{F}_j \, , 
\end{equation}
for $i = 1,2,3$, wherein $\boldsymbol{\mu}_{ij}$ stands for the hydrodynamic mobility tensor relating between the translational velocity of sphere~$i$ and the force exerted on sphere~$j$.
The hydrodynamic mobility incorporates the effect of the many-body fluid-mediated interactions between suspended particles.
Here, we confine ourselves for simplicity to the situation in which only contributions stemming from self $(i=j)$ and pair $(i \ne j)$ hydrodynamic interactions are accounted for. 
We will assess the accuracy of our approach with direct comparison with fully resolved numerical simulations based on the finite element method (Sec.~\ref{sec:FEM}).

In the so-called point-particle approximation, in which $R \ll R_\mathrm{c}$, the scaled self-mobility function is given to leading order in~$R/R_\mathrm{c}$ by~\cite{bohlin1960drag, daddi2017hydrodynamic}
\begin{equation}
    \frac{\mu_{ii}}{\mu_0} = 1 + \delta \, \frac{R}{R_\mathrm{c}} \, , 
    \label{eq:self-mobility}
\end{equation}
wherein $\mu_0 = 1/ \left( 6\pi\eta R \right)$ is the bulk mobility, and
\begin{equation}
    \delta = -\frac{3}{2\pi} 
    \int_0^\infty \frac{A(s)}{B(s)} \, \mathrm{d} s \, . 
\label{eq:self_int}
\end{equation}
Here, we have defined
\begin{subequations}
\begin{align}
    A(s) &= 4 I_1(s) K_0(s) + s^2 \big( I_0 (s) K_1(s) + I_1(s) K_0(s) \big) \notag \\
    &\quad- 2 s \big( I_0(s) K_0(s) + I_1(s) K_1(s) \big) \, , \\ 
    B(s) &=  2 I_0(s) I_1(s) + s \big( I_1(s)^2 - I_0(s)^2 \big) \, , 
\end{align}
\end{subequations}
with $I_\nu$ and~$K_\nu$ denoting the $\nu$-th~order modified Bessel functions (also known as the hyperbolic Bessel functions) of the first and second kinds, respectively.
A numerical evaluation of the infinite integral in Eq.~\eqref{eq:self_int} yields
\begin{equation}
       \delta \simeq -2.10444 \, . 
\label{eq:self}
\end{equation}

Analogously, the hydrodynamic pair mobility in the point-particle approximation is given in a scaled form by~\cite{daddi2017hydrodynamic, daddi2017diffusion}
\begin{equation}
    \frac{\mu_{ij}}{\mu_0} = \frac{3}{2} \frac{R}{R_\mathrm{c}} \left( \frac{1}{\sigma} + \xi_{ij} (\sigma) \right) \, , 
    \label{eq:pair-mobility}
\end{equation}
where
\begin{equation}
    \xi_{ij} (\sigma) = -\frac{1}{\pi} 
    \int_0^\infty \frac{A(s)}{B(s)} \, \cos \left( \sigma s \right) \, \mathrm{d} s \, , 
\label{eq:pair_int}
\end{equation}
wherein $\sigma = \left|(\mathbf{r}_i - \mathbf{r}_j) \cdot \hat{\mathbf{e}}_z\right|/R_\mathrm{c}$.
Clearly, $\mu_{ij} = \mu_{ji}$, as required by symmetry.
In particular, $\delta/\xi(0) = 3/2$.

It is worth noting that the pair mobility can likewise be expressed in terms of converging infinite series of the form~\cite{cui2002screened}
\begin{equation}
      \frac{\mu_{ij}}{\mu_0} =  
      \frac{3}{4} \sum_{n=1}^{\infty} \varphi_n e^{-\alpha_n \sigma} \, , \label{eq:pair_series}
\end{equation}
where
\begin{equation}
    \varphi_n = a_n \cos(\beta_n \sigma) + b_n \sin (\beta_n \sigma) \, .
\end{equation}
Here, $u_n := \alpha_n + i \beta_n$ are the complex roots of the equation
\begin{equation}
    u_n (J_0^2 (u_n) + J_1^2(u_n)) = 2 J_0(u_n) J_1 (u_n) = 0 \, .
\end{equation}
In addition,
\begin{align}
    a_n + i b_n &= 2 \Big(\pi \big( 2J_1(u_n) Y_0(u_n) - u_n (J_0(u_n) Y_0(u_n) \notag \\
    &\quad \left. + \, J_1(u_n) Y_1(u_n)) \big) - u_n \Big) \middle / J_1^2(u_n) \right. \, , 
\end{align}
where $J_\nu$ and $Y_\nu$ stand for the $\nu$-th order Bessel functions of the first and second kinds, respectively.
Accordingly, the pair mobility function displays a sharp exponential decay as the distance between particles becomes larger.
In the limit $\sigma  \gg 1$, the series in Eq.~\eqref{eq:pair_series} can be truncated to the first term to yield
\begin{equation}
\frac{\mu_{ij}}{\mu_0} \simeq \frac{3}{4} \big( a_1 \cos(\beta_1 \sigma) + b_1 \sin (\beta_1 \sigma) \big) e^{-\alpha_1 \sigma} \, , 
\end{equation}
with the numerical estimates $\alpha_1 \simeq 4.46630$, $\beta_1 \simeq 1.46747$, $a_1 \simeq -0.03698$ and $b_1 \simeq 13.80821$.

Differentiating Eq.~\eqref{eq:gh} with respect to time yields $V_2 = V_1 + \dot{g}$ and~$V_3 = V_1 - \dot{h}$, with dots standing for a time derivative and $V_i =  \mathbf{V}_i \cdot \hat{\mathbf{e}}_z$ denotes the axial velocity along the centerline of the confining tube.
By requiring the force-free condition (Eq.~\ref{eqn:Forcefree}), we find that the instantaneous axial velocity of the center sphere is obtained as
\begin{equation}
    V_1 = \frac{\dot{h} \left( \mu_{ii}-\mu_{12} \right) M_+ - \dot{g} \left( \mu_{ii}-\mu_{13} \right) M_-}{3\mu_{ii}^2 - 2\mu_{ii} \left( \mu_{12} + \mu_{13} + \mu_{23} \right) - N} \, , 
    \label{eq:V1_final}
\end{equation}
wherein $M_\pm = \mu_{ii} \pm \mu_{12} \mp \mu_{13} - \mu_{23}$ and $N = \mu_{12}^2 + \left( \mu_{13} - \mu_{23} \right)^2 - 2 \mu_{12} \left( \mu_{13} + \mu_{23} \right)$.
We have $\big( \dot{g}, \dot{h} \big) = \left(0,-W \right)$ for $t \in [0,T/4]$,
$\big( \dot{g}, \dot{h} \big) = \left(-W, 0 \right)$ for $t \in [T/4, T/2]$,
$\big( \dot{g}, \dot{h} \big) = \left(0,W \right)$ for $t \in [T/2, 3T/4]$,
and $\big( \dot{g}, \dot{h} \big) = \left(W,0 \right)$ for $t \in [3T/4, T]$.
We recall that $\mu_{12} = \mu_{ij} \left( \sigma = g/R_\mathrm{c} \right) $, $\mu_{13} = \mu_{ij} \left( \sigma = h/R_\mathrm{c} \right) $, and $\mu_{23} = \mu_{ij} \left( \sigma = (g+h)/R_\mathrm{c} \right)$.
We note that self, $\mu_{ii}$, and pair, $\mu_{ij}$, mobilities are given by Eqs.~\eqref{eq:self-mobility} and~\eqref{eq:pair-mobility}, respectively.

Finally, the mean swimming velocity is obtained by averaging over one full cycle as
\begin{equation}
    \overline{V}_1 = \frac{1}{T} \int_0^T V_1(t) \, \mathrm{d} t \, . 
\end{equation}

Owing to the delicate and peculiar nature of the resulting axial speed stated by Eq.~\eqref{eq:V1_final}, an analytical evaluation of the mean is rather complicated and far from being trivial, even in the simplistic situation without confinement.
To be able to make analytical progress, we expand perturbatively the axial velocity in the small parameter $R/D$.
By substituting the expressions of the self- and pair-mobility functions, given by Eqs.~\eqref{eq:self-mobility} and~\eqref{eq:pair-mobility}, respectively, into Eq.~\eqref{eq:V1_final}, and noting that $N = \mathcal{O}\left( \left( R/D \right)^2 \right)$, the instantaneous swimming velocity can readily, upon Taylor expansion in the small parameter $R/D$, be cast in the form
\begin{equation}
    V_1 = V_1^\mathrm{B} + V_1^\mathrm{C} + \mathcal{O} \left( \left( \tfrac{R}{D} \right)^2 \right) \, , 
\end{equation}
where $V_1^\mathrm{B}$ in the instantaneous velocity in the absence of confinement, given by
\begin{align}
    V_1^\mathrm{B} &= \Big( \dot{g} \big( (R-2g)h^2 - 2(R+h)g^2 \big) 
    +  \dot{h} \big( (2h-R)g^2 \notag \\
    &\quad+ \left. 2(R+g)h^2 \big) \Big) \middle/ 6gh(g+h) \right. \, .
\end{align}
Moreover, $V_1^\mathrm{C}$ is the confinement-related contribution to the instantaneous velocity, given by
\begin{equation}
    V_1^\mathrm{C} = \frac{R}{6R_\mathrm{c}} \left(\dot{g} \, \Xi_1 - \dot{h} \, \Xi_2 \right) \, , 
\end{equation}
wherein $\Xi_1 = \xi_{12}-2\xi_{13} + \xi_{23}$ and $\Xi_2 = \xi_{13} - 2\xi_{12} + \xi_{23}$.

We find that the bulk-related contribution to the average speed is obtained as
\begin{equation}
    \overline{V}_1^\mathrm{B} = \frac{R}{3T} \left( \frac{2\epsilon^2}{D(D-\epsilon)} 
    + \ln \left( \frac{4D(D-\epsilon)}{\left( 2D-\epsilon \right)^2} \right) \right) \, . \label{eqn:Bulk}
\end{equation}
In particular, for $\epsilon \ll D$, we get
\begin{align}
\overline{V}_1^\mathrm{B} = \frac{7R}{12 T} \left( \left(\frac{\epsilon}{D} \right)^2+\left(\frac{\epsilon}{D} \right)^3  \right) + \mathcal{O} \left( \left(\frac{\epsilon}{D} \right)^4 \right). \label{eqn:BulkAsym}
\end{align}

\begin{figure} 
	\centering
	\includegraphics[width=0.4\textwidth]{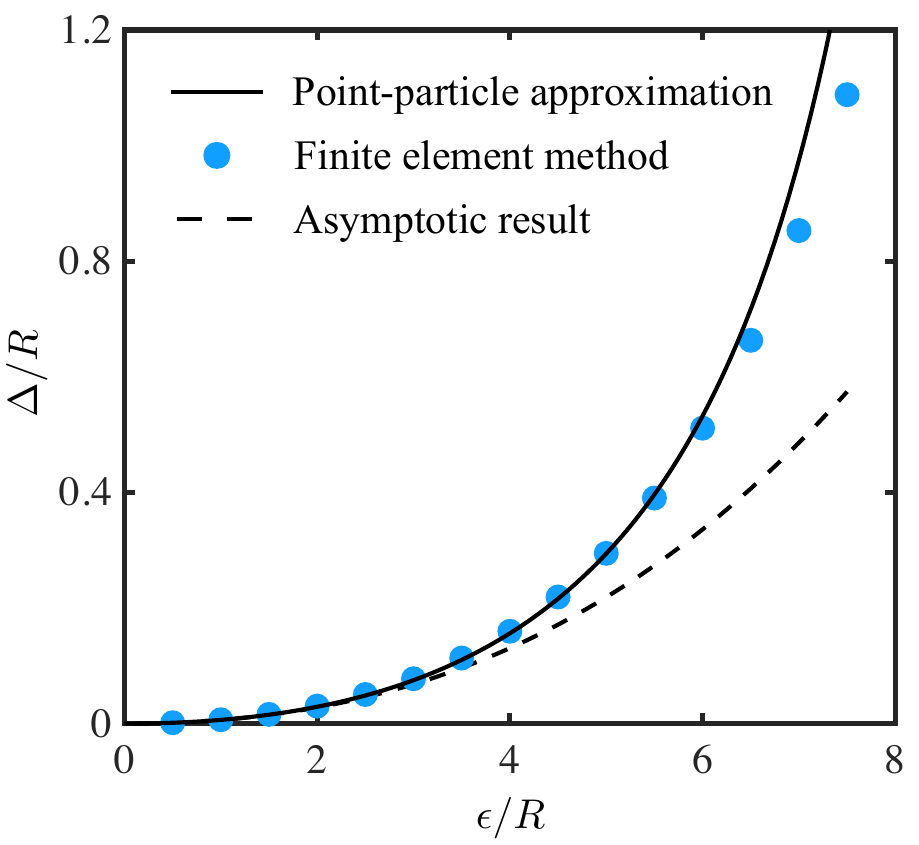}
	\caption{The scaled net displacement of the swimmer per cycle, $\Delta/R$, in an unbounded fluid as a function of the scaled contraction length of the swimmer, $\epsilon/R$. The black solid line represents results based on the point-particle approximation, $\Delta =  \overline{V}_1^\mathrm{B} T$, where $ \overline{V}_1^\mathrm{B}$ is given by Eq.~\ref{eqn:Bulk}, whereas the black dashed line corresponds to the asymptotic results given by Eq.~\ref{eqn:BulkAsym} (or Eq.~\ref{eqn:AsymDisp}) in the limit $\epsilon \ll D$. The blue circles are simulation results based on the finite element method with a large radius of confinement $R_c/R=1000$ to simulate an unbounded fluid domain. Here $D/R=10$.
 } 
	\label{fig2}
\end{figure}

The contribution to the averaged speed due to confinement can be approximated in the limit $R \ll D$ as
\begin{equation}
    \overline{V}_1^\mathrm{C} = \frac{R}{3\pi T} \int_0^\infty
    \frac{A(s)}{B(s)} \left( \frac{2\epsilon}{R_\mathrm{c}} \, \psi_1(s) - \frac{\psi_2(s)}{s} \right) \mathrm{d} s  \, , 
    \label{eq:V1C}
\end{equation}
where
\begin{align}
    \psi_1 (s) &= \cos \left( \tfrac{D}{R_\mathrm{c}} \, s \right) - 
    \cos \left( \tfrac{D-\epsilon}{R_\mathrm{c}} \, s \right) , \notag \\
    \psi_2 (s) &= \sin \left( \tfrac{2D}{R_\mathrm{c}} \, s \right)
    - 2 \sin \left( \tfrac{2D-\epsilon}{R_\mathrm{c}} \, s \right)
    + \sin \left( \tfrac{2 \left( D-\epsilon\right)}{R_\mathrm{c}} \, s \right) . \notag 
\end{align}
Here, we have swapped the order of integration with respect to~$s$ and~$t$. 
It is worth highlighting that Eq.~\eqref{eq:V1C}, which provides the confinement-related contribution to the averaged swimming speed, remains valid across the entire range of values for $D$ and $R_\mathrm{c}$. The only assumption made to derive the approximate expressions for the swimming speed is that $R$ is significantly smaller than $D$.
Since Eq.~\eqref{eq:V1C} involves infinite integrals over the scaled wavenumber~$s$, corresponding analytical expressions cannot be obtained in the limit $\epsilon \ll D$, unlike the case for the bulk-related contribution given by Eq.~\eqref{eqn:BulkAsym}.

\subsection{Finite element method} \label{sec:FEM}
We also perform fully coupled numerical simulations of the momentum (Eq.~\ref{eqn:Stokes1}) and continuity (Eq.~\ref{eqn:Stokes2}) equations using the finite element method (FEM) implemented in the COMSOL Multiphysics environment. We compare these numerical simulation results, which capture the full sphere-sphere and sphere-confinement hydrodynamic interactions, with predictions based on the point-particle approximation in Sec.~\ref{sec:point-particle}. The axisymmetry of the problem setup reduces the computational complexity of the problem from three-dimensional to two-dimensional. Since Stokes flows have slow spatial decay, in order to minimize any hydrodynamic influence from the ends, we consider cylindrical computational domain of radius $R_c$ and a long axial length of approximately 2000$R$ ($1000R$ in each direction away from the outer spheres). The domain is discretized by about 20,000--35,000 P3--P2 (third-order for fluid velocity and second-order for pressure) triangular mesh elements, with local mesh refinement in the proximity of the three spheres. The degree of freedom is of the order of (0.5$-$1)$~\times 10^6$, depending on the radius of the confining tube. We use the Multifrontal Massively Parallel Sparse (MUMPS) direct solver for all simulations.

Due to the time independence of Stokes flows, the motion of the swimmer is completely determined by its instantaneous movement and geometrical configuration. To simulate the swimming motion over a full cycle, the movement of the swimmer is broken down into separate, stationary simulations for different time instants in individual strokes. At each instant, the velocities of the spheres are determined by the relative motion of the three spheres plus an unknown swimming speed in the axial direction on all spheres. These prescribed velocities on the spheres are implemented as boundary conditions on the spheres. To determine the unknown swimming speed, the force-free condition (Eq.~\ref{eqn:Forcefree}) is implemented as a global equation, which is solved together with the momentum and continuity equations to obtain the swimming speed, velocity field, and pressure field simultaneously at each time instant in the swimming cycle. We then perform numerical integration of the swimming speed over a full cycle to obtain the net displacement of the swimmer per cycle, $\Delta$.

\begin{figure} 
	\centering
	\includegraphics[width=0.423\textwidth]{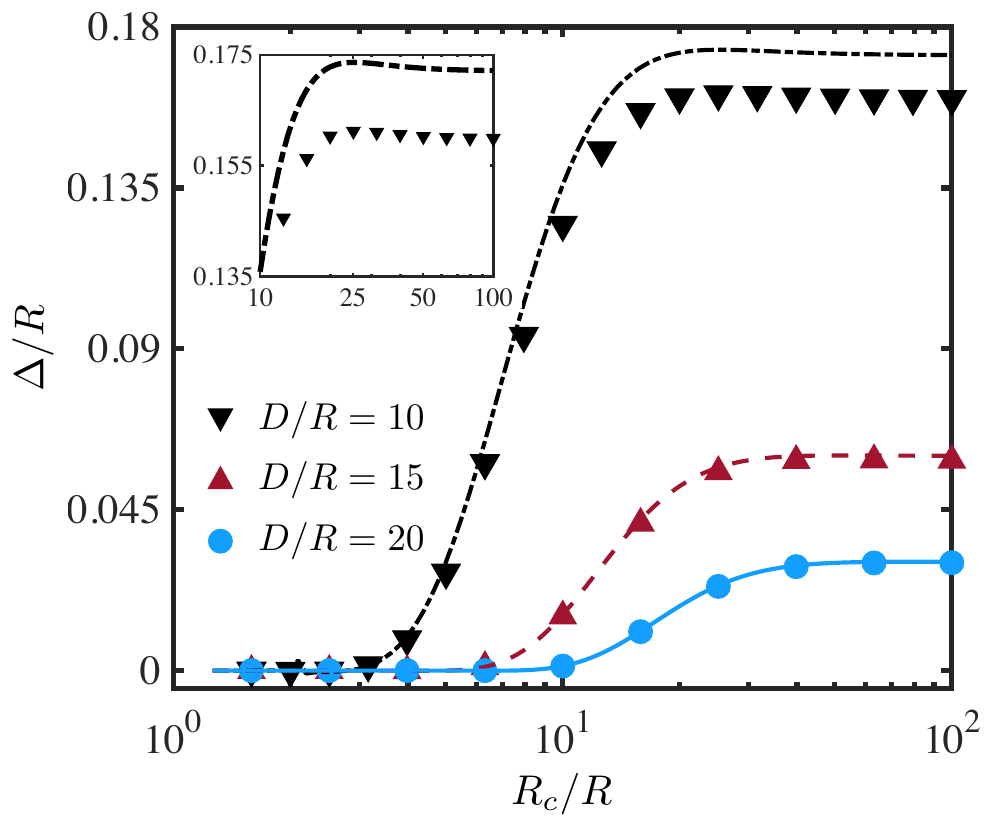}
	\caption{The scaled net displacement of the swimmer per cycle, $\Delta/R$, as a function of the scaled capillary tube radius, $R_c/R$, for different values of scaled rod lengths, $D/R$. Here $\epsilon/R=4$. The symbols represent numerical results from FEM simulations (see legends); the lines with the same colors represent the corresponding predictions from the point-particle approximation by integrating the instantaneous velocity given by Eq.~\ref{eq:V1_final} over one full cycle. Inset: a magnified view of the non-monotonic variation of the scaled net displacement with the scaled capillary radius for $D/R=10$.}
	\label{fig3}
\end{figure}

\section{Results and discussion} \label{sec: Results and discussion}

In the following sections, we first cross-validate the point-particle approximation and numerical simulations based on the finite element method by considering the motion of a three-sphere swimmer in an unbounded fluid domain in Sec.~\ref{sec:unbounded}. We then characterize in Sec.~\ref{sec:Confinement} the effect of axisymmetric confinement on the propulsion performance of the three-sphere swimmer for different levels of confinement and properties of the swimmer.

\begin{figure} 
	\centering
	\includegraphics[width=0.4\textwidth]{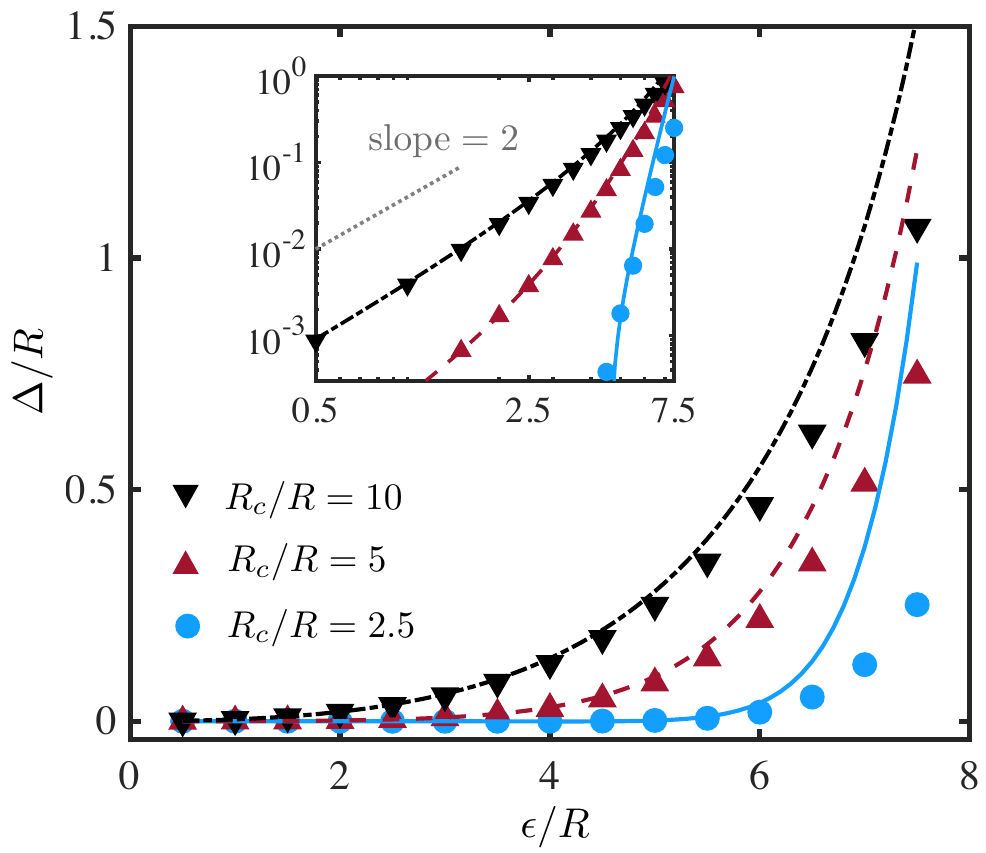}
	\caption{The scaled net displacement of the swimmer per cycle, $\Delta/R$, as a function of the scaled contraction length of the swimmer, $\epsilon/R$, for different values of the scaled capillary tube radius, $R_c/R$. Here $D/R=10$. The symbols represent numerical results from FEM simulations (see legends); the lines with the same colors represent the corresponding predictions from the point-particle approximation by integrating the instantaneous velocity given by Eq.~\ref{eq:V1_final} over one full cycle. The inset displays a log-log plot of the results, where a dotted grey line of slope 2 is added to aid visualization.} 
	\label{fig4}
\end{figure}

\subsection{Swimming in an unbounded fluid} \label{sec:unbounded}

\begin{figure*} 
	\centering
	\includegraphics[width=1\textwidth]{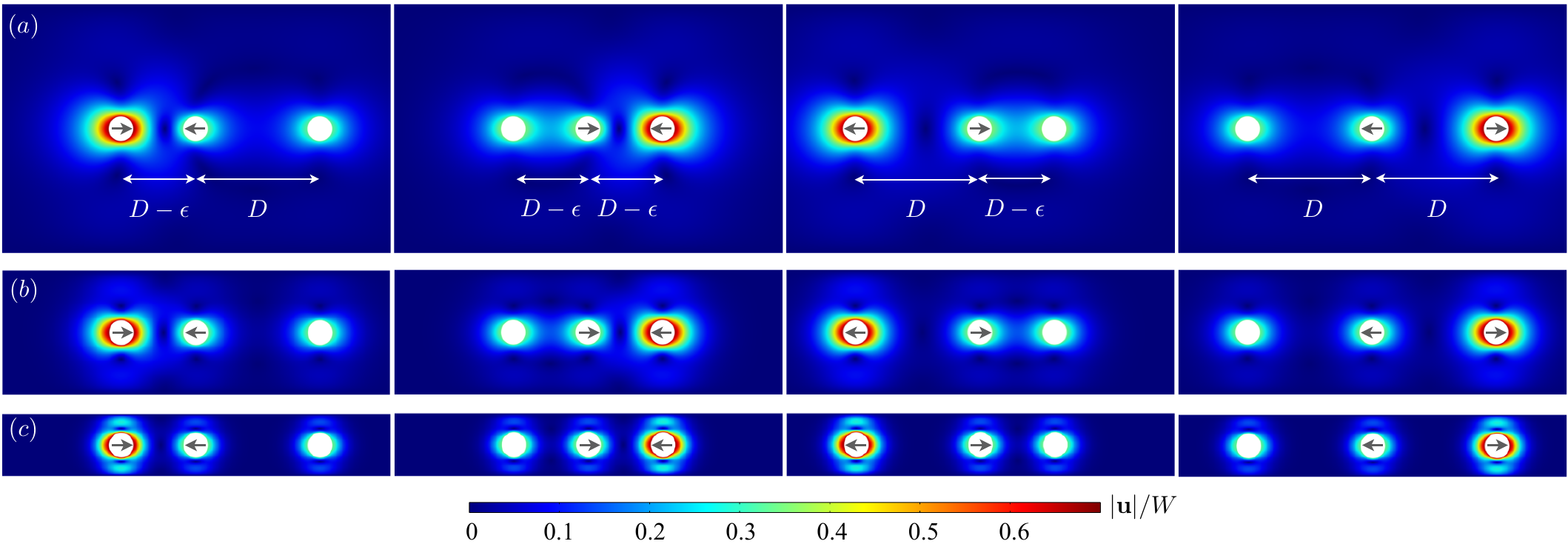}
	\caption{The scaled flow speed distribution, $|\mathbf{u}|/W$, around the swimmer at the end of individual strokes I--IV (from left to right panels) illustrated in Fig.~\ref{fig1} for different values of the scaled capillary tube radius: (a) $R_c/R=10$, (b) $R_c/R=5$, and (c) $R_c/R=2.5$. The black arrows indicate the relative motion (contraction or extension) of the pair of spheres in different strokes. Here $D/R=10$ and $\epsilon/R=4$.} 
 \label{fig5}
\end{figure*}

For validation, we consider the motion of the three-sphere swimmer in an unbounded fluid domain. In the particle-particle approximation (Sec.~\ref{sec:point-particle}), the bulk-related contribution to the average speed $\overline{V}^B_1$ is given by Eq.~\ref{eqn:Bulk}, which can be multiplied by the period $T$ to obtain the net displacement of the swimmer per cycle, $\Delta$, shown in Fig.~\ref{fig2} (black solid line). When $\epsilon \ll D$, the net displacement is calculated from Eq.~\ref{eqn:BulkAsym} as 
\begin{align}
\Delta \sim \frac{7R}{12} \left( \left(\frac{\epsilon}{D} \right)^2+\left(\frac{\epsilon}{D} \right)^3  \right), \label{eqn:AsymDisp}
\end{align}
which is represented by the black-dashed line in Fig.~\ref{fig2}. We remark that the asymptotic result given in Eq.~\ref{eqn:AsymDisp} is consistent with that given by Earl \textit{et al.} \cite{Earl2007}, which rectified the result presented in Najafi and Golestanian \cite{Najafi2004}. The asymptotic result in Eq.~\ref{eqn:AsymDisp} reveals that the net displacement of the swimmer per cycle scales quadratically with the contraction length of the swimmer, $\Delta=\mathcal{O}(\epsilon^2)$, in the regime of $\epsilon/D \ll 1$. 

To compare with the above theoretical results, in the FEM simulations we use an exceedingly large radius of confinement ($R_c/R = 1000$) to simulate the swimming motion in an unbounded fluid domain.
The FEM results are represented by blue circles in Fig.~\ref{fig2}. The comparison between theoretical and numerical results show that the point-particle approximation captures quantitatively the propulsion behaviors for small to moderate contraction lengths of the swimmer, where the spheres are sufficiently distanced from each other throughout the swimming cycle. For larger contraction lengths, the spheres come closer to each other during contraction, leading to more significant hydrodynamic interactions between the spheres. Consequently, the point-particle approximation starts to deviate from the FEM results, over-estimating the net displacement of the swimmer \cite{Earl2007,Paz2023}. Despite these deviations, the point-particle approximation continues to capture the qualitative trend of the propulsion behavior.

\subsection{Swimming under axisymmetric confinement} \label{sec:Confinement}

In Fig.~\ref{fig3}, we probe the effect of axisymmetric confinement by examining the net displacement of the three-sphere swimmer along the centerline of a capillary tube. Here we keep the contraction length constant ($\epsilon/R = 4$) and only vary the radius of the tube, $R_c$. For a given fully extended arm length $D$, the effect of confinement becomes significant when $R_c/R$ is of $\mathcal{O}(1)-\mathcal{O}(10)$: in this regime, the results reveal that a tighter confinement (decreasing $R_c$) substantially reduces the net displacement of the swimmer. We note that this trend is in stark contrast with helical propulsion in a capillary tube \cite{Liu2014}, where the confinement largely enhances propulsion, illustrating how confinement can have qualitatively distinct effects on swimming depending on the underlying propulsion mechanisms. Results from the FEM simulations (symbols) and point-particle approximation (lines) agree well when the spheres are separated by large arm lengths (e.g., $D/R=15, \ 20$), when the sphere-sphere and sphere-confinement hydrodynamic interactions are expected to be weaker. For $D/R=10$, the point-particle approximation still captures properly the qualitative feature of the confined swimming motion when the spheres are in closer proximity, where the near-field effects become more pronounced. As a remark, while the major effect here is a substantial and rapid decay in propulsion under tight confinements, the swimmer also displays a very slight enhancement in propulsion when the confining radius is large (e.g., when $R_c/R$ is beyond $O(10)$ for $D/R=10$). This minute effect, captured by both the FEM simulations and point-particle approximation, is observed for all values of $D/R$ presented in Fig.~\ref{fig3} and is more apparent for the case of $D/R=10$ (inset).

Next, we probe how the net displacement of the swimmer varies with its contraction length when swimming in a capillary tube (Fig.~\ref{fig4}). While the net displacement of the swimmer grows with the contraction length in general, it occurs at different rates depending on the degree of confinement (\textit{i.e}, the value of $R_c/R$). In an unbounded fluid, the net displacement grows quadratically with the contraction length, $\Delta/R \sim 7 \epsilon^2/(12D)$, as given by Eq.~\ref{eqn:AsymDisp}. In Fig.~\ref{fig4} inset, we consider a log-log plot of the results to better visualize the scaling. For a relatively loose confinement $R_c/R = 10$ (downside black triangles and black dot-dash line), the log-log plot shows an approximately quadratic scaling between the net displacement and the contraction length, similar to the case in an unbounded fluid. However, as the environment becomes more confined ($R_c/R=2.5$ and $5$), results from both point-particle approximation (red dashed line and blue solid line) and FEM (red upside triangles and blue circles) indicate slopes increasingly greater than two in the inset. These results illustrate that the scaling goes beyond second-order in confined swimming; the three-sphere mechanism becomes increasingly ineffective in generating a net displacement under tighter confinement.

To develop a more physical understanding of the above results, we revisit the symmetry arguments by Najafi and Golestanian \cite{Najafi2004} that showed how the four strokes in the cycle are related. These arguments remain valid for the swimmer under axisymmetric confinement considered in this work: stroke III is related to stroke II upon a left-right reflection and a time-reversal transformations, whereas stroke IV is related to stroke I with the same transformations. Consequently, the net displacement of the swimmer after executing a full cycle is simply reduced to (two times) the difference in the net displacement of the center sphere generated by stroke I and stroke II. Stroke I generates a net displacement to the left, while stroke II generates a net displacement to the right. It is crucial to note that these net displacements differ in their magnitudes because the force acting on the spheres when they are far apart (in stroke I) is different from when they are in closer proximity (in stroke II) due to interactions between the spheres via their surrounding flows. The hydrodynamic interactions lead to only partial cancellation of the displacements generated by strokes I and II, giving rise to the net displacement of the swimmer after a cycle. When the hydrodynamic interaction is neglected, the two strokes would generate displacements with equal magnitudes in opposite directions, cancelling each other and yielding zero net propulsion.

Based on the above understanding of the propulsion mechanism, we attribute the reduced net displacement of the confined swimmer to weakened hydrodynamic interactions among the spheres under confinement as follows. It was shown that the flow due to a Stokeslet decays exponentially in a capillary tube due to the confinement \cite{liron_shahar_1978}, as opposed to decaying as the inverse of the distance in an unbounded fluid. The flow around the moving spheres of the swimmer in a capillary tube is therefore expected to decay more rapidly in space. 
To visualize this effect, we plot in Fig.~\ref{fig5} the flow field surrounding the swimmer at different time instants in a swimming cycle with different levels of confinement. As the radius of the confining tube decreases from $R_c/R=10$ in panel (a) to $R_c/R=2.5$ in panel (c), the magnitude of the flow around individual spheres can be observed to decay more rapidly away from the spheres. These faster spatial decays of the flow velocity weaken the hydrodynamic interaction between the spheres, thereby reducing the hydrodynamic difference between the case when the spheres are more far apart in stroke I and the case when they are in closer proximity in stroke II. The reduced hydrodynamic difference between the two strokes therefore generates displacements with more similar magnitudes, leading to the reduced net displacements of the swimmer in a capillary tube as observed in Fig.~\ref{fig3}.

\section{Concluding Remarks} \label{sec:conclusion}

In this work, we examine the propulsion of a three-sphere swimmer along the centerline of a capillary tube at low Re. We combine theoretical analysis via the point-particle approximation and simulations based on the finite element method to uncover how the propulsion speed varies with the radius of the confining tube as well as geometric and kinematic properties of the swimmer. The results show that the presence of confinement does not significantly affect the propulsion speed until the scaled radius of the confining tube is in the range of $R_c/R = \mathcal{O}(1)-\mathcal{O}(10)$, where the swimmer exhibits sharp decays in propulsion speed as the radius of the tube decreases. The presence of confinement also leads to higher-order scaling between the net displacement and the contraction length of the swimmer, reducing the effectiveness of this propulsion mechanism. We contrast the reduced propulsion speed observed here with the enhanced helical propulsion inside a capillary tube reported earlier \cite{Liu2014}, highlighting how the effect of confinement can manifest in qualitatively different manners depending on the swimmer's propulsion mechanism. While helical propulsion is based on the drag anisotropy of slender bodies, the three-sphere swimmer here relies on the sphere-sphere hydrodynamic interactions--a physically different mechanism--to self-propel. The reduced propulsion performance observed here is attributed to the more rapid spatial decays of the flow velocity of moving bodies in a tube, which reduces the hydrodynamic interaction between the spheres and thereby the net displacement of the swimmer.   

Based on the above physical understanding of the results, we hypothesize that the propulsion of a three-sphere swimmer in porous media may also be hindered due to the screening of hydrodynamic interactions by networks of obstacles, in contrast to enhanced propulsion predicted for  different types of swimmers in heterogeneous viscous environments \cite{Leshansky2009,Fu_2010,Nguyenho2016,Leiderman2016,PhysRevFluids.3.094102, PhysRevResearch.5.033030, hosaka2023hydrodynamics}. An investigation is currently underway to evaluate this hypothesis and will be reported in a future work. Furthermore, we considered the effect due to rigid confinement in this work, while the fluid-structure interaction between the swimmer and elastic confinements can have profound impacts on the swimming performance \cite{LedesmaAguilar2013,daddi2019frequency,Dalal2020}. It would be worthwhile to consider the case of an elastic tube and systematically examine the interplay between shear and bending deformation modes in prescribing the hydrodynamics of the swimmer under elastic confinement. Finally, we focus on the effect of axisymmetric confinement here to preserve the one-dimensional nature of the motion, which allows us to measure how the degree of confinement affects the propulsion speed in a simple manner. Lifting this restriction to examine more general motion of a three-sphere swimmer in a capillary tube could lead to more complex and interesting swimming dynamics in future studies.\\

 \begin{acknowledgments}
O.S.P. acknowledges funding support by the National Science Foundation (Grant No.~1830958). A.D.-M.-I. acknowledges support from the Max Planck Center Twente for Complex Fluid Dynamics, the Max Planck School Matter to Life, and the MaxSynBio Consortium, which are jointly funded by the Federal Ministry of Education and Research (BMBF) of Germany and the Max Planck Society. We are also grateful for the computational resources from the WAVE computing facility (enabled by the E.~L.~Wiegand Foundation) at Santa Clara University.
 \end{acknowledgments}

 \appendix 

 \section{A confined three-sphere swimmer with harmonic oscillations of the rod lengths} \label{sec:Appendix}
\begin{figure} 
	\centering
	\includegraphics[width=0.43\textwidth]{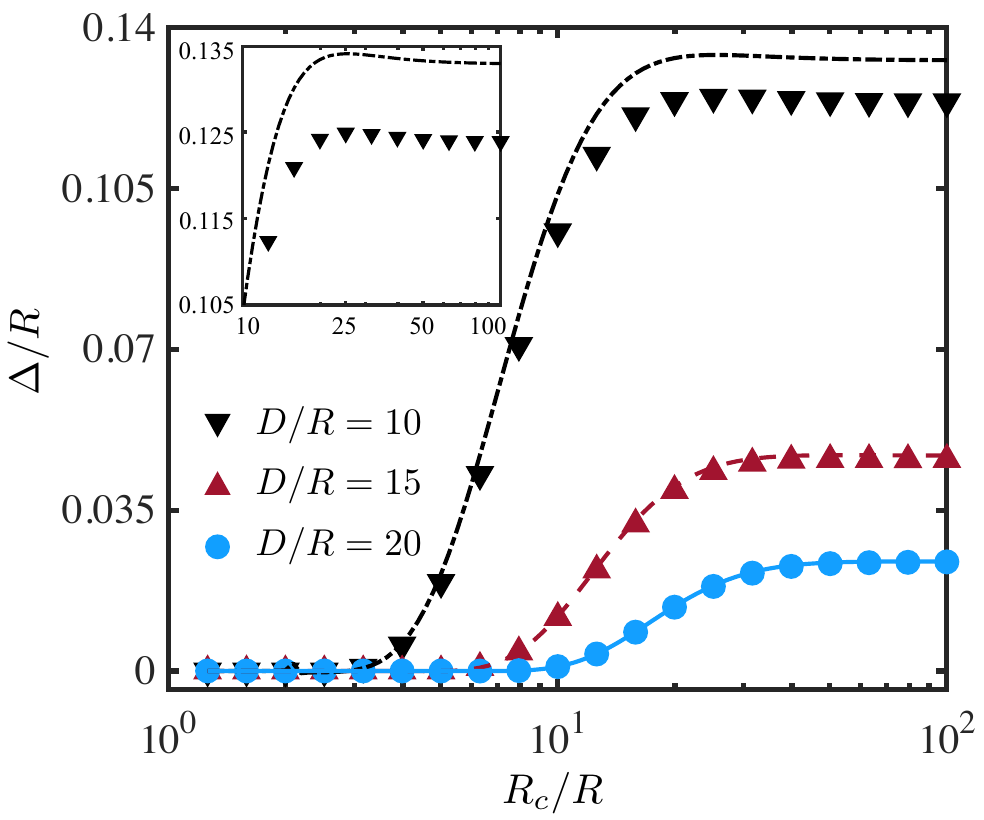}
	\caption{The scaled net displacement of the swimmer per cycle, $\Delta/R$, as a function of the scaled capillary tube radius, $R_c/R$, for different values of scaled rod lengths, $D/R$, for the case of harmonic deformations prescribed by Eqs.~\ref{eqn:HarmonicDef1} and \ref{eqn:HarmonicDef2}. Here $\epsilon/R=4$. The symbols represent numerical results from FEM simulations (see legends); the lines with the same colors represent the corresponding predictions from the point-particle approximation by integrating the instantaneous velocity given by Eq.~\ref{eq:V1_final} over one full cycle.  Inset: a magnified view of the non-monotonic variation of the scaled net displacement with the scaled capillary radius for $D/R=10$.}
	\label{App}
\end{figure}

\begin{figure} 
	\centering
	\includegraphics[width=0.43\textwidth]{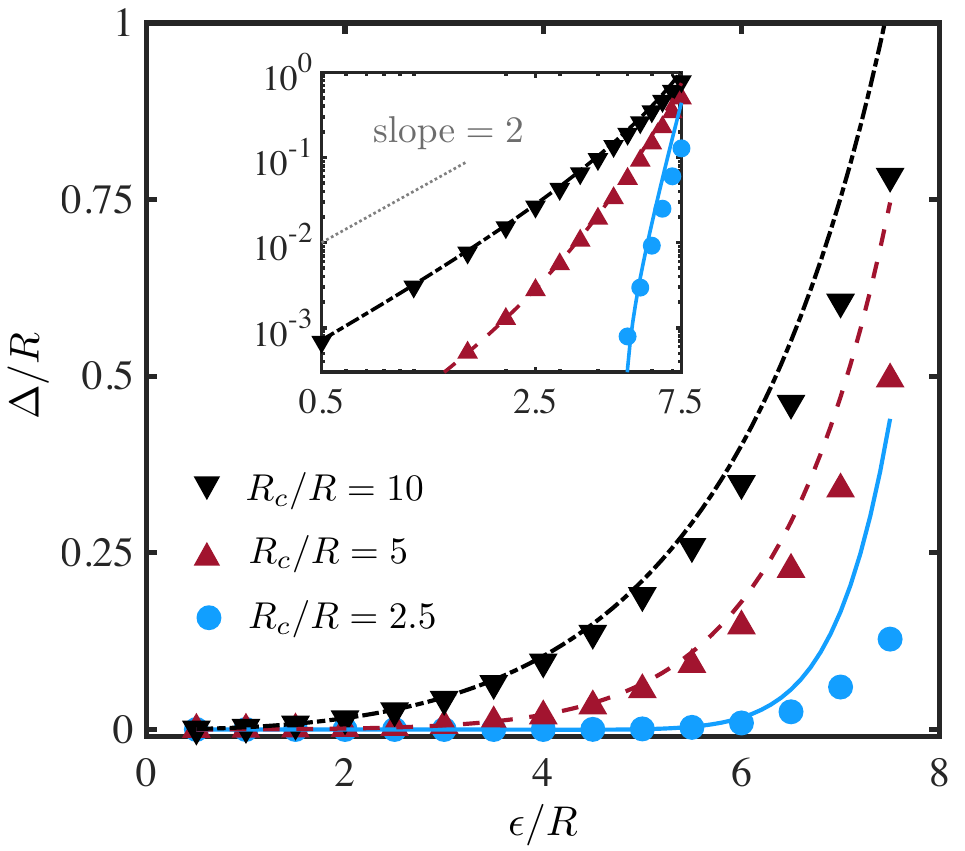}
	\caption{The scaled net displacement of the swimmer per cycle, $\Delta/R$, as a function of the scaled contraction length of the swimmer, $\epsilon/R$, for different values of the scaled capillary tube radius, $R_c/R$, for the case of harmonic deformations prescribed by Eqs.~\ref{eqn:HarmonicDef1} and \ref{eqn:HarmonicDef2}. Here $D/R=10$. The symbols represent numerical results from FEM simulations (see legends); the lines with the same colors represent the corresponding predictions from the point-particle approximation by integrating the instantaneous velocity given by Eq.~\ref{eq:V1_final} over one full cycle. The inset displays a log-log plot of the results, where a dotted gray line of slope 2 is added to aid visualization.} 
	\label{App_Eps}
\end{figure}

We follow Najafi and Golestanian \cite{Najafi2004} in the main text in considering a constant relative speed $W$ in the change of arm length. In this appendix, we consider also harmonic deformations of the arms \cite{GolestanianAjdari2008} to establish some generality of the  conclusion. Specifically, we prescribe the following variations, respectively, for the length of front and rear rods,
\begin{align}
g(t) &= D-\frac{\epsilon}{2}+\frac{\epsilon}{2}\cos (\omega t),\label{eqn:HarmonicDef1}\\
h(t) &= D-\frac{\epsilon}{2}+\frac{\epsilon}{2}\cos (\omega t+\phi). \label{eqn:HarmonicDef2}
\end{align}
The two rods have an equilibrium length of $D-\epsilon/2$ with sinusoidal oscillations of amplitude $\epsilon/2$, angular frequency $\omega$, and a phase mismatch $\phi$. Here we set $\omega = \pi W/\epsilon$, so that the period of oscillation is given by $T=2\epsilon/W$. As a remark, when $\phi=0$, the swimmer generates zero net propulsion by symmetry; when $\phi=\pi$, the overall deformation of the swimmer becomes reciprocal motion, which also leads to zero net propulsion as dictated by the scallop theorem. Here we present results for the specific case of $\phi=\pi/2$, which was shown to generate the maximum amount of net displacement of the swimmer in an unbounded fluid \cite{GolestanianAjdari2008}. As shown in Figs.~\ref{App} and \ref{App_Eps}, a confined swimmer with harmonic variations of its arm lengths exhibit qualitatively the same behaviors, compared with the case of a constant rate of change of the arm lengths examined in the main text (Figs.~\ref{fig3} and \ref{fig4}).

 \section*{Data Availability}
The data that support the findings of this study are available from the corresponding author upon reasonable request.

%


\begin{thebibliography}{90}%
\makeatletter
\providecommand \@ifxundefined [1]{%
 \@ifx{#1\undefined}
}%
\providecommand \@ifnum [1]{%
 \ifnum #1\expandafter \@firstoftwo
 \else \expandafter \@secondoftwo
 \fi
}%
\providecommand \@ifx [1]{%
 \ifx #1\expandafter \@firstoftwo
 \else \expandafter \@secondoftwo
 \fi
}%
\providecommand \natexlab [1]{#1}%
\providecommand \enquote  [1]{``#1''}%
\providecommand \bibnamefont  [1]{#1}%
\providecommand \bibfnamefont [1]{#1}%
\providecommand \citenamefont [1]{#1}%
\providecommand \href@noop [0]{\@secondoftwo}%
\providecommand \href [0]{\begingroup \@sanitize@url \@href}%
\providecommand \@href[1]{\@@startlink{#1}\@@href}%
\providecommand \@@href[1]{\endgroup#1\@@endlink}%
\providecommand \@sanitize@url [0]{\catcode `\\12\catcode `\$12\catcode
  `\&12\catcode `\#12\catcode `\^12\catcode `\_12\catcode `\%12\relax}%
\providecommand \@@startlink[1]{}%
\providecommand \@@endlink[0]{}%
\providecommand \url  [0]{\begingroup\@sanitize@url \@url }%
\providecommand \@url [1]{\endgroup\@href {#1}{\urlprefix }}%
\providecommand \urlprefix  [0]{URL }%
\providecommand \Eprint [0]{\href }%
\providecommand \doibase [0]{https://doi.org/}%
\providecommand \selectlanguage [0]{\@gobble}%
\providecommand \bibinfo  [0]{\@secondoftwo}%
\providecommand \bibfield  [0]{\@secondoftwo}%
\providecommand \translation [1]{[#1]}%
\providecommand \BibitemOpen [0]{}%
\providecommand \bibitemStop [0]{}%
\providecommand \bibitemNoStop [0]{.\EOS\space}%
\providecommand \EOS [0]{\spacefactor3000\relax}%
\providecommand \BibitemShut  [1]{\csname bibitem#1\endcsname}%
\let\auto@bib@innerbib\@empty
\bibitem [{\citenamefont {Brennen}\ and\ \citenamefont
  {Winet}(1977)}]{Brennen1977}%
  \BibitemOpen
  \bibfield  {author} {\bibinfo {author} {\bibfnamefont {C.}~\bibnamefont
  {Brennen}}\ and\ \bibinfo {author} {\bibfnamefont {H.}~\bibnamefont
  {Winet}},\ }\bibfield  {title} {\enquote {\bibinfo {title} {Fluid mechanics
  of propulsion by cilia and flagella},}\ }\href@noop {} {\bibfield  {journal}
  {\bibinfo  {journal} {Annu. Rev. Fluid Mech.}\ }\textbf {\bibinfo {volume}
  {9}},\ \bibinfo {pages} {339--398} (\bibinfo {year} {1977})}\BibitemShut
  {NoStop}%
\bibitem [{\citenamefont {Bray}()}]{Bray2000}%
  \BibitemOpen
  \bibfield  {author} {\bibinfo {author} {\bibfnamefont {D.}~\bibnamefont
  {Bray}},\ }\href@noop {} {\emph {\bibinfo {title} {{Cell Movements}}}}\
  (\bibinfo  {publisher} {New York, Garland, U.S.A.})\BibitemShut {NoStop}%
\bibitem [{\citenamefont {Lauga}(2020)}]{EricLauga2020}%
  \BibitemOpen
  \bibfield  {author} {\bibinfo {author} {\bibfnamefont {E.}~\bibnamefont
  {Lauga}},\ }\href@noop {} {\emph {\bibinfo {title} {The Fluid Dynamics of
  Cell Motility}}},\ Vol.~\bibinfo {volume} {62}\ (\bibinfo  {publisher}
  {Cambridge University Press, Cambridge, U.K.},\ \bibinfo {year}
  {2020})\BibitemShut {NoStop}%
\bibitem [{\citenamefont {Nelson}, \citenamefont {Kaliakatsos},\ and\
  \citenamefont {Abbott}(2010)}]{Nelson2010}%
  \BibitemOpen
  \bibfield  {author} {\bibinfo {author} {\bibfnamefont {B.~J.}\ \bibnamefont
  {Nelson}}, \bibinfo {author} {\bibfnamefont {I.~K.}\ \bibnamefont
  {Kaliakatsos}},\ and\ \bibinfo {author} {\bibfnamefont {J.~J.}\ \bibnamefont
  {Abbott}},\ }\bibfield  {title} {\enquote {\bibinfo {title} {Microrobots for
  minimally invasive medicine},}\ }\href@noop {} {\bibfield  {journal}
  {\bibinfo  {journal} {Annu. Rev. Biomed. Eng.}\ }\textbf {\bibinfo {volume}
  {12}},\ \bibinfo {pages} {55--85} (\bibinfo {year} {2010})}\BibitemShut
  {NoStop}%
\bibitem [{\citenamefont {{Sitti}}\ \emph {et~al.}(2015)\citenamefont
  {{Sitti}}, \citenamefont {{Ceylan}}, \citenamefont {{Hu}}, \citenamefont
  {{Giltinan}}, \citenamefont {{Turan}}, \citenamefont {{Yim}},\ and\
  \citenamefont {{Diller}}}]{Sitti2015}%
  \BibitemOpen
  \bibfield  {author} {\bibinfo {author} {\bibfnamefont {M.}~\bibnamefont
  {{Sitti}}}, \bibinfo {author} {\bibfnamefont {H.}~\bibnamefont {{Ceylan}}},
  \bibinfo {author} {\bibfnamefont {W.}~\bibnamefont {{Hu}}}, \bibinfo {author}
  {\bibfnamefont {J.}~\bibnamefont {{Giltinan}}}, \bibinfo {author}
  {\bibfnamefont {M.}~\bibnamefont {{Turan}}}, \bibinfo {author} {\bibfnamefont
  {S.}~\bibnamefont {{Yim}}},\ and\ \bibinfo {author} {\bibfnamefont
  {E.}~\bibnamefont {{Diller}}},\ }\bibfield  {title} {\enquote {\bibinfo
  {title} {Biomedical applications of untethered mobile milli/microrobots},}\
  }\href@noop {} {\bibfield  {journal} {\bibinfo  {journal} {Proc. IEEE Inst.
  Electr. Electron Eng.}\ }\textbf {\bibinfo {volume} {103}},\ \bibinfo {pages}
  {205--224} (\bibinfo {year} {2015})}\BibitemShut {NoStop}%
\bibitem [{\citenamefont {Li}\ \emph {et~al.}(2017)\citenamefont {Li},
  \citenamefont {Esteban-Fernández~de Ávila}, \citenamefont {Gao},
  \citenamefont {Zhang},\ and\ \citenamefont {Wang}}]{JinxingLi2017}%
  \BibitemOpen
  \bibfield  {author} {\bibinfo {author} {\bibfnamefont {J.}~\bibnamefont
  {Li}}, \bibinfo {author} {\bibfnamefont {B.}~\bibnamefont
  {Esteban-Fernández~de Ávila}}, \bibinfo {author} {\bibfnamefont
  {W.}~\bibnamefont {Gao}}, \bibinfo {author} {\bibfnamefont {L.}~\bibnamefont
  {Zhang}},\ and\ \bibinfo {author} {\bibfnamefont {J.}~\bibnamefont {Wang}},\
  }\bibfield  {title} {\enquote {\bibinfo {title} {Micro/nanorobots for
  biomedicine: Delivery, surgery, sensing, and detoxification},}\ }\href@noop
  {} {\bibfield  {journal} {\bibinfo  {journal} {Sci. Robot.}\ }\textbf
  {\bibinfo {volume} {2}},\ \bibinfo {pages} {eaam6431} (\bibinfo {year}
  {2017})}\BibitemShut {NoStop}%
\bibitem [{\citenamefont {Hu}\ \emph {et~al.}(2018)\citenamefont {Hu},
  \citenamefont {Lum}, \citenamefont {Mastrangeli},\ and\ \citenamefont
  {Sitti}}]{WenqiHu2018}%
  \BibitemOpen
  \bibfield  {author} {\bibinfo {author} {\bibfnamefont {W.}~\bibnamefont
  {Hu}}, \bibinfo {author} {\bibfnamefont {G.~Z.}\ \bibnamefont {Lum}},
  \bibinfo {author} {\bibfnamefont {M.}~\bibnamefont {Mastrangeli}},\ and\
  \bibinfo {author} {\bibfnamefont {M.}~\bibnamefont {Sitti}},\ }\bibfield
  {title} {\enquote {\bibinfo {title} {Small-scale soft-bodied robot with
  multimodal locomotion},}\ }\href@noop {} {\bibfield  {journal} {\bibinfo
  {journal} {Nature}\ }\textbf {\bibinfo {volume} {554}},\ \bibinfo {pages}
  {81--85} (\bibinfo {year} {2018})}\BibitemShut {NoStop}%
\bibitem [{\citenamefont {Palagi}\ and\ \citenamefont
  {Fischer}(2018)}]{Palagi2018}%
  \BibitemOpen
  \bibfield  {author} {\bibinfo {author} {\bibfnamefont {S.}~\bibnamefont
  {Palagi}}\ and\ \bibinfo {author} {\bibfnamefont {P.}~\bibnamefont
  {Fischer}},\ }\bibfield  {title} {\enquote {\bibinfo {title} {Bioinspired
  microrobots},}\ }\href@noop {} {\bibfield  {journal} {\bibinfo  {journal}
  {Nat. Rev. Mater.}\ }\textbf {\bibinfo {volume} {3}},\ \bibinfo {pages}
  {113--124} (\bibinfo {year} {2018})}\BibitemShut {NoStop}%
\bibitem [{\citenamefont {Tsang}\ \emph
  {et~al.}(2020{\natexlab{a}})\citenamefont {Tsang}, \citenamefont {Demir},
  \citenamefont {Ding},\ and\ \citenamefont {Pak}}]{Tsang2020}%
  \BibitemOpen
  \bibfield  {author} {\bibinfo {author} {\bibfnamefont {A.~C.}\ \bibnamefont
  {Tsang}}, \bibinfo {author} {\bibfnamefont {E.}~\bibnamefont {Demir}},
  \bibinfo {author} {\bibfnamefont {Y.}~\bibnamefont {Ding}},\ and\ \bibinfo
  {author} {\bibfnamefont {O.~S.}\ \bibnamefont {Pak}},\ }\bibfield  {title}
  {\enquote {\bibinfo {title} {Roads to smart artificial microswimmers},}\
  }\href@noop {} {\bibfield  {journal} {\bibinfo  {journal} {Adv. Intell.
  Syst.}\ }\textbf {\bibinfo {volume} {2}},\ \bibinfo {pages} {1900137}
  (\bibinfo {year} {2020}{\natexlab{a}})}\BibitemShut {NoStop}%
\bibitem [{\citenamefont {Purcell}(1977)}]{Purcell1977}%
  \BibitemOpen
  \bibfield  {author} {\bibinfo {author} {\bibfnamefont {E.~M.}\ \bibnamefont
  {Purcell}},\ }\bibfield  {title} {\enquote {\bibinfo {title} {{Life at low
  Reynolds number}},}\ }\href@noop {} {\bibfield  {journal} {\bibinfo
  {journal} {Am. J. Phys.}\ }\textbf {\bibinfo {volume} {45}},\ \bibinfo
  {pages} {3--11} (\bibinfo {year} {1977})}\BibitemShut {NoStop}%
\bibitem [{\citenamefont {Fauci}\ and\ \citenamefont {Dillon}(2006)}]{Fauci06}%
  \BibitemOpen
  \bibfield  {author} {\bibinfo {author} {\bibfnamefont {L.~J.}\ \bibnamefont
  {Fauci}}\ and\ \bibinfo {author} {\bibfnamefont {R.}~\bibnamefont {Dillon}},\
  }\bibfield  {title} {\enquote {\bibinfo {title} {Biofluidmechanics of
  reproduction},}\ }\href@noop {} {\bibfield  {journal} {\bibinfo  {journal}
  {Annu. Rev. Fluid Mech.}\ }\textbf {\bibinfo {volume} {38}},\ \bibinfo
  {pages} {371--394} (\bibinfo {year} {2006})}\BibitemShut {NoStop}%
\bibitem [{\citenamefont {Lauga}\ and\ \citenamefont
  {Powers}(2009)}]{LaugaPowers2009}%
  \BibitemOpen
  \bibfield  {author} {\bibinfo {author} {\bibfnamefont {E.}~\bibnamefont
  {Lauga}}\ and\ \bibinfo {author} {\bibfnamefont {T.~R.}\ \bibnamefont
  {Powers}},\ }\bibfield  {title} {\enquote {\bibinfo {title} {The
  hydrodynamics of swimming microorganisms},}\ }\href@noop {} {\bibfield
  {journal} {\bibinfo  {journal} {Rep. Prog. Phys.}\ }\textbf {\bibinfo
  {volume} {72}},\ \bibinfo {pages} {096601} (\bibinfo {year}
  {2009})}\BibitemShut {NoStop}%
\bibitem [{\citenamefont {Yeomans}, \citenamefont {Pushkin},\ and\
  \citenamefont {Shum}(2014)}]{Yeomans2014}%
  \BibitemOpen
  \bibfield  {author} {\bibinfo {author} {\bibfnamefont {J.~M.}\ \bibnamefont
  {Yeomans}}, \bibinfo {author} {\bibfnamefont {D.~O.}\ \bibnamefont
  {Pushkin}},\ and\ \bibinfo {author} {\bibfnamefont {H.}~\bibnamefont
  {Shum}},\ }\bibfield  {title} {\enquote {\bibinfo {title} {An introduction to
  the hydrodynamics of swimming microorganisms},}\ }\href@noop {} {\bibfield
  {journal} {\bibinfo  {journal} {Eur. Phys. J. Spec. Top.}\ }\textbf {\bibinfo
  {volume} {223}},\ \bibinfo {pages} {1771--1785} (\bibinfo {year}
  {2014})}\BibitemShut {NoStop}%
\bibitem [{\citenamefont {Bechinger}\ \emph {et~al.}(2016)\citenamefont
  {Bechinger}, \citenamefont {Di~Leonardo}, \citenamefont {L{\"o}wen},
  \citenamefont {Reichhardt}, \citenamefont {Volpe},\ and\ \citenamefont
  {Volpe}}]{bechinger2016active}%
  \BibitemOpen
  \bibfield  {author} {\bibinfo {author} {\bibfnamefont {C.}~\bibnamefont
  {Bechinger}}, \bibinfo {author} {\bibfnamefont {R.}~\bibnamefont
  {Di~Leonardo}}, \bibinfo {author} {\bibfnamefont {H.}~\bibnamefont
  {L{\"o}wen}}, \bibinfo {author} {\bibfnamefont {C.}~\bibnamefont
  {Reichhardt}}, \bibinfo {author} {\bibfnamefont {G.}~\bibnamefont {Volpe}},\
  and\ \bibinfo {author} {\bibfnamefont {G.}~\bibnamefont {Volpe}},\ }\bibfield
   {title} {\enquote {\bibinfo {title} {Active particles in complex and crowded
  environments},}\ }\href@noop {} {\bibfield  {journal} {\bibinfo  {journal}
  {Rev. Mod. Phys.}\ }\textbf {\bibinfo {volume} {88}},\ \bibinfo {pages}
  {045006} (\bibinfo {year} {2016})}\BibitemShut {NoStop}%
\bibitem [{\citenamefont {Lauga}(2016)}]{Lauga2016}%
  \BibitemOpen
  \bibfield  {author} {\bibinfo {author} {\bibfnamefont {E.}~\bibnamefont
  {Lauga}},\ }\bibfield  {title} {\enquote {\bibinfo {title} {Bacterial
  hydrodynamics},}\ }\href@noop {} {\bibfield  {journal} {\bibinfo  {journal}
  {Annu. Rev. Fluid Mech.}\ }\textbf {\bibinfo {volume} {48}},\ \bibinfo
  {pages} {105--130} (\bibinfo {year} {2016})}\BibitemShut {NoStop}%
\bibitem [{\citenamefont {Wan}(2022)}]{Wan2022}%
  \BibitemOpen
  \bibfield  {author} {\bibinfo {author} {\bibfnamefont {K.}~\bibnamefont
  {Wan}},\ }\bibfield  {title} {\enquote {\bibinfo {title} {The beat of
  isolated cilia.}}\ }\href@noop {} {\bibfield  {journal} {\bibinfo  {journal}
  {Nat. Phys.}\ }\textbf {\bibinfo {volume} {18}},\ \bibinfo {pages} {234--235}
  (\bibinfo {year} {2022})}\BibitemShut {NoStop}%
\bibitem [{\citenamefont {Abbott}\ \emph {et~al.}(2009)\citenamefont {Abbott},
  \citenamefont {Peyer}, \citenamefont {Lagomarsino}, \citenamefont {Zhang},
  \citenamefont {Dong}, \citenamefont {Kaliakatsos},\ and\ \citenamefont
  {Nelson}}]{Abbott2009}%
  \BibitemOpen
  \bibfield  {author} {\bibinfo {author} {\bibfnamefont {J.~J.}\ \bibnamefont
  {Abbott}}, \bibinfo {author} {\bibfnamefont {K.~E.}\ \bibnamefont {Peyer}},
  \bibinfo {author} {\bibfnamefont {M.~C.}\ \bibnamefont {Lagomarsino}},
  \bibinfo {author} {\bibfnamefont {L.}~\bibnamefont {Zhang}}, \bibinfo
  {author} {\bibfnamefont {L.}~\bibnamefont {Dong}}, \bibinfo {author}
  {\bibfnamefont {I.~K.}\ \bibnamefont {Kaliakatsos}},\ and\ \bibinfo {author}
  {\bibfnamefont {B.~J.}\ \bibnamefont {Nelson}},\ }\bibfield  {title}
  {\enquote {\bibinfo {title} {How should microrobots swim?}}\ }\href@noop {}
  {\bibfield  {journal} {\bibinfo  {journal} {Int. J. Robotics Res.}\ }\textbf
  {\bibinfo {volume} {28}},\ \bibinfo {pages} {1434--1447} (\bibinfo {year}
  {2009})}\BibitemShut {NoStop}%
\bibitem [{\citenamefont {Ebbens}\ and\ \citenamefont
  {Howse}(2010)}]{Ebbens2010}%
  \BibitemOpen
  \bibfield  {author} {\bibinfo {author} {\bibfnamefont {S.~J.}\ \bibnamefont
  {Ebbens}}\ and\ \bibinfo {author} {\bibfnamefont {J.~R.}\ \bibnamefont
  {Howse}},\ }\bibfield  {title} {\enquote {\bibinfo {title} {In pursuit of
  propulsion at the nanoscale},}\ }\href@noop {} {\bibfield  {journal}
  {\bibinfo  {journal} {Soft Matter}\ }\textbf {\bibinfo {volume} {6}},\
  \bibinfo {pages} {726--738} (\bibinfo {year} {2010})}\BibitemShut {NoStop}%
\bibitem [{\citenamefont {Lauga}(2011)}]{Lauga2011}%
  \BibitemOpen
  \bibfield  {author} {\bibinfo {author} {\bibfnamefont {E.}~\bibnamefont
  {Lauga}},\ }\bibfield  {title} {\enquote {\bibinfo {title} {Life around the
  scallop theorem},}\ }\href@noop {} {\bibfield  {journal} {\bibinfo  {journal}
  {Soft Matter}\ }\textbf {\bibinfo {volume} {7}},\ \bibinfo {pages}
  {3060--3065} (\bibinfo {year} {2011})}\BibitemShut {NoStop}%
\bibitem [{\citenamefont {Sharan}\ \emph {et~al.}(2023)\citenamefont {Sharan},
  \citenamefont {Daddi-Moussa-Ider}, \citenamefont {Agudo-Canalejo},
  \citenamefont {Golestanian},\ and\ \citenamefont
  {Simmchen}}]{sharan2023pair}%
  \BibitemOpen
  \bibfield  {author} {\bibinfo {author} {\bibfnamefont {P.}~\bibnamefont
  {Sharan}}, \bibinfo {author} {\bibfnamefont {A.}~\bibnamefont
  {Daddi-Moussa-Ider}}, \bibinfo {author} {\bibfnamefont {J.}~\bibnamefont
  {Agudo-Canalejo}}, \bibinfo {author} {\bibfnamefont {R.}~\bibnamefont
  {Golestanian}},\ and\ \bibinfo {author} {\bibfnamefont {J.}~\bibnamefont
  {Simmchen}},\ }\bibfield  {title} {\enquote {\bibinfo {title} {Pair
  interaction between two catalytically active colloids},}\ }\href@noop {}
  {\bibfield  {journal} {\bibinfo  {journal} {Small}\ ,\ \bibinfo {pages}
  {2300817}} (\bibinfo {year} {2023})}\BibitemShut {NoStop}%
\bibitem [{\citenamefont {Becker}, \citenamefont {Koehler},\ and\ \citenamefont
  {Stone}(2003)}]{becker2003self}%
  \BibitemOpen
  \bibfield  {author} {\bibinfo {author} {\bibfnamefont {L.~E.}\ \bibnamefont
  {Becker}}, \bibinfo {author} {\bibfnamefont {S.~A.}\ \bibnamefont
  {Koehler}},\ and\ \bibinfo {author} {\bibfnamefont {H.~A.}\ \bibnamefont
  {Stone}},\ }\bibfield  {title} {\enquote {\bibinfo {title} {{On
  self-propulsion of micro-machines at low Reynolds number: Purcell's
  three-link swimmer}},}\ }\href@noop {} {\bibfield  {journal} {\bibinfo
  {journal} {J. Fluid Mech.}\ }\textbf {\bibinfo {volume} {490}},\ \bibinfo
  {pages} {15--35} (\bibinfo {year} {2003})}\BibitemShut {NoStop}%
\bibitem [{\citenamefont {Tam}\ and\ \citenamefont
  {Hosoi}(2007)}]{tam2007optimal}%
  \BibitemOpen
  \bibfield  {author} {\bibinfo {author} {\bibfnamefont {D.}~\bibnamefont
  {Tam}}\ and\ \bibinfo {author} {\bibfnamefont {A.~E.}\ \bibnamefont
  {Hosoi}},\ }\bibfield  {title} {\enquote {\bibinfo {title} {Optimal stroke
  patterns for {P}urcell's three-link swimmer},}\ }\href@noop {} {\bibfield
  {journal} {\bibinfo  {journal} {Phys. Rev. Lett.}\ }\textbf {\bibinfo
  {volume} {98}},\ \bibinfo {pages} {068105} (\bibinfo {year}
  {2007})}\BibitemShut {NoStop}%
\bibitem [{\citenamefont {Avron}\ and\ \citenamefont {Raz}(2008)}]{Avron2008}%
  \BibitemOpen
  \bibfield  {author} {\bibinfo {author} {\bibfnamefont {J.~E.}\ \bibnamefont
  {Avron}}\ and\ \bibinfo {author} {\bibfnamefont {O.}~\bibnamefont {Raz}},\
  }\bibfield  {title} {\enquote {\bibinfo {title} {A geometric theory of
  swimming: Purcell's swimmer and its symmetrized cousin},}\ }\href@noop {}
  {\bibfield  {journal} {\bibinfo  {journal} {New J. Phys.}\ }\textbf {\bibinfo
  {volume} {10}},\ \bibinfo {pages} {063016} (\bibinfo {year}
  {2008})}\BibitemShut {NoStop}%
\bibitem [{\citenamefont {Qin}\ \emph {et~al.}(2023)\citenamefont {Qin},
  \citenamefont {Zou}, \citenamefont {Zhu},\ and\ \citenamefont
  {Pak}}]{Qin2023_B}%
  \BibitemOpen
  \bibfield  {author} {\bibinfo {author} {\bibfnamefont {K.}~\bibnamefont
  {Qin}}, \bibinfo {author} {\bibfnamefont {Z.}~\bibnamefont {Zou}}, \bibinfo
  {author} {\bibfnamefont {L.}~\bibnamefont {Zhu}},\ and\ \bibinfo {author}
  {\bibfnamefont {O.~S.}\ \bibnamefont {Pak}},\ }\bibfield  {title} {\enquote
  {\bibinfo {title} {{Reinforcement learning of a multi-link swimmer at low
  Reynolds numbers}},}\ }\href@noop {} {\bibfield  {journal} {\bibinfo
  {journal} {Phys. Fluids}\ }\textbf {\bibinfo {volume} {35}},\ \bibinfo
  {pages} {032003} (\bibinfo {year} {2023})}\BibitemShut {NoStop}%
\bibitem [{\citenamefont {Qin}\ and\ \citenamefont {Pak}(2023)}]{Qin2023}%
  \BibitemOpen
  \bibfield  {author} {\bibinfo {author} {\bibfnamefont {K.}~\bibnamefont
  {Qin}}\ and\ \bibinfo {author} {\bibfnamefont {O.~S.}\ \bibnamefont {Pak}},\
  }\bibfield  {title} {\enquote {\bibinfo {title} {Purcell's swimmer in a
  shear-thinning fluid},}\ }\href@noop {} {\bibfield  {journal} {\bibinfo
  {journal} {Phys. Rev. Fluids}\ }\textbf {\bibinfo {volume} {8}},\ \bibinfo
  {pages} {033301} (\bibinfo {year} {2023})}\BibitemShut {NoStop}%
\bibitem [{\citenamefont {Najafi}\ and\ \citenamefont
  {Golestanian}(2004)}]{Najafi2004}%
  \BibitemOpen
  \bibfield  {author} {\bibinfo {author} {\bibfnamefont {A.}~\bibnamefont
  {Najafi}}\ and\ \bibinfo {author} {\bibfnamefont {R.}~\bibnamefont
  {Golestanian}},\ }\bibfield  {title} {\enquote {\bibinfo {title} {{Simple
  swimmer at low Reynolds number: Three linked spheres}},}\ }\href@noop {}
  {\bibfield  {journal} {\bibinfo  {journal} {Phys. Rev. E}\ }\textbf {\bibinfo
  {volume} {69}},\ \bibinfo {pages} {062901} (\bibinfo {year}
  {2004})}\BibitemShut {NoStop}%
\bibitem [{\citenamefont {Avron}, \citenamefont {Kenneth},\ and\ \citenamefont
  {Oaknin}(2005)}]{Avron2005}%
  \BibitemOpen
  \bibfield  {author} {\bibinfo {author} {\bibfnamefont {J.~E.}\ \bibnamefont
  {Avron}}, \bibinfo {author} {\bibfnamefont {O.}~\bibnamefont {Kenneth}},\
  and\ \bibinfo {author} {\bibfnamefont {D.~H.}\ \bibnamefont {Oaknin}},\
  }\bibfield  {title} {\enquote {\bibinfo {title} {Pushmepullyou: an efficient
  micro-swimmer},}\ }\href@noop {} {\bibfield  {journal} {\bibinfo  {journal}
  {New J. Phys.}\ }\textbf {\bibinfo {volume} {7}},\ \bibinfo {pages} {234}
  (\bibinfo {year} {2005})}\BibitemShut {NoStop}%
\bibitem [{\citenamefont {Earl}\ \emph {et~al.}(2007)\citenamefont {Earl},
  \citenamefont {Pooley}, \citenamefont {Ryder}, \citenamefont {Bredberg},\
  and\ \citenamefont {Yeomans}}]{Earl2007}%
  \BibitemOpen
  \bibfield  {author} {\bibinfo {author} {\bibfnamefont {D.~J.}\ \bibnamefont
  {Earl}}, \bibinfo {author} {\bibfnamefont {C.~M.}\ \bibnamefont {Pooley}},
  \bibinfo {author} {\bibfnamefont {J.~F.}\ \bibnamefont {Ryder}}, \bibinfo
  {author} {\bibfnamefont {I.}~\bibnamefont {Bredberg}},\ and\ \bibinfo
  {author} {\bibfnamefont {J.~M.}\ \bibnamefont {Yeomans}},\ }\bibfield
  {title} {\enquote {\bibinfo {title} {{Modeling microscopic swimmers at low
  Reynolds number}},}\ }\href@noop {} {\bibfield  {journal} {\bibinfo
  {journal} {J. Chem. Phys.}\ }\textbf {\bibinfo {volume} {126}},\ \bibinfo
  {pages} {064703} (\bibinfo {year} {2007})}\BibitemShut {NoStop}%
\bibitem [{\citenamefont {Golestanian}\ and\ \citenamefont
  {Ajdari}(2009)}]{Golestanian2009}%
  \BibitemOpen
  \bibfield  {author} {\bibinfo {author} {\bibfnamefont {R.}~\bibnamefont
  {Golestanian}}\ and\ \bibinfo {author} {\bibfnamefont {A.}~\bibnamefont
  {Ajdari}},\ }\bibfield  {title} {\enquote {\bibinfo {title} {{Stochastic low
  Reynolds number swimmers}},}\ }\href@noop {} {\bibfield  {journal} {\bibinfo
  {journal} {J. Phys. Condens. Matter}\ }\textbf {\bibinfo {volume} {21}},\
  \bibinfo {pages} {204104} (\bibinfo {year} {2009})}\BibitemShut {NoStop}%
\bibitem [{\citenamefont {Alouges}, \citenamefont {DeSimone},\ and\
  \citenamefont {Lefebvre}(2008)}]{Alouges2008}%
  \BibitemOpen
  \bibfield  {author} {\bibinfo {author} {\bibfnamefont {F.}~\bibnamefont
  {Alouges}}, \bibinfo {author} {\bibfnamefont {A.}~\bibnamefont {DeSimone}},\
  and\ \bibinfo {author} {\bibfnamefont {A.}~\bibnamefont {Lefebvre}},\
  }\bibfield  {title} {\enquote {\bibinfo {title} {Optimal strokes for low
  {R}eynolds number swimmers: An example},}\ }\href@noop {} {\bibfield
  {journal} {\bibinfo  {journal} {J. Nonlinear Sci.}\ }\textbf {\bibinfo
  {volume} {18}},\ \bibinfo {pages} {277--302} (\bibinfo {year}
  {2008})}\BibitemShut {NoStop}%
\bibitem [{\citenamefont {Alouges}\ \emph {et~al.}(2012)\citenamefont
  {Alouges}, \citenamefont {DeSimone}, \citenamefont {Heltai}, \citenamefont
  {Lefebvre-Lepot},\ and\ \citenamefont {Merlet}}]{Alouges2013B}%
  \BibitemOpen
  \bibfield  {author} {\bibinfo {author} {\bibfnamefont {F.}~\bibnamefont
  {Alouges}}, \bibinfo {author} {\bibfnamefont {A.}~\bibnamefont {DeSimone}},
  \bibinfo {author} {\bibfnamefont {L.}~\bibnamefont {Heltai}}, \bibinfo
  {author} {\bibfnamefont {A.}~\bibnamefont {Lefebvre-Lepot}},\ and\ \bibinfo
  {author} {\bibfnamefont {B.}~\bibnamefont {Merlet}},\ }\bibfield  {title}
  {\enquote {\bibinfo {title} {Optimally swimming {S}tokesian robots},}\
  }\href@noop {} {\bibfield  {journal} {\bibinfo  {journal} {Discrete
  Continuous Dyn. Syst. Ser. B}\ }\textbf {\bibinfo {volume} {18}},\ \bibinfo
  {pages} {1189} (\bibinfo {year} {2012})}\BibitemShut {NoStop}%
\bibitem [{\citenamefont {Montino}\ and\ \citenamefont
  {DeSimone}(2015)}]{Montino2015}%
  \BibitemOpen
  \bibfield  {author} {\bibinfo {author} {\bibfnamefont {A.}~\bibnamefont
  {Montino}}\ and\ \bibinfo {author} {\bibfnamefont {A.}~\bibnamefont
  {DeSimone}},\ }\bibfield  {title} {\enquote {\bibinfo {title} {{Three-sphere
  low-Reynolds-number swimmer with a passive elastic arm}},}\ }\href@noop {}
  {\bibfield  {journal} {\bibinfo  {journal} {Eur. Phys. J. E}\ }\textbf
  {\bibinfo {volume} {38}},\ \bibinfo {pages} {42} (\bibinfo {year}
  {2015})}\BibitemShut {NoStop}%
\bibitem [{\citenamefont {Wang}\ and\ \citenamefont {Othmer}(2018)}]{Wang2018}%
  \BibitemOpen
  \bibfield  {author} {\bibinfo {author} {\bibfnamefont {Q.}~\bibnamefont
  {Wang}}\ and\ \bibinfo {author} {\bibfnamefont {H.~G.}\ \bibnamefont
  {Othmer}},\ }\bibfield  {title} {\enquote {\bibinfo {title} {Analysis of a
  model microswimmer with applications to blebbing cells and mini-robots},}\
  }\href@noop {} {\bibfield  {journal} {\bibinfo  {journal} {J. Math. Biol.}\
  }\textbf {\bibinfo {volume} {76}},\ \bibinfo {pages} {1699--1763} (\bibinfo
  {year} {2018})}\BibitemShut {NoStop}%
\bibitem [{\citenamefont {Wang}(2019)}]{Wang2019}%
  \BibitemOpen
  \bibfield  {author} {\bibinfo {author} {\bibfnamefont {Q.}~\bibnamefont
  {Wang}},\ }\bibfield  {title} {\enquote {\bibinfo {title} {Optimal strokes of
  low {Reynolds} number linked-sphere swimmers},}\ }\href@noop {} {\bibfield
  {journal} {\bibinfo  {journal} {Appl. Sci.}\ }\textbf {\bibinfo {volume}
  {9}},\ \bibinfo {pages} {4023} (\bibinfo {year} {2019})}\BibitemShut
  {NoStop}%
\bibitem [{\citenamefont {Nasouri}, \citenamefont {Vilfan},\ and\ \citenamefont
  {Golestanian}(2019)}]{Nasouri2019}%
  \BibitemOpen
  \bibfield  {author} {\bibinfo {author} {\bibfnamefont {B.}~\bibnamefont
  {Nasouri}}, \bibinfo {author} {\bibfnamefont {A.}~\bibnamefont {Vilfan}},\
  and\ \bibinfo {author} {\bibfnamefont {R.}~\bibnamefont {Golestanian}},\
  }\bibfield  {title} {\enquote {\bibinfo {title} {Efficiency limits of the
  three-sphere swimmer},}\ }\href@noop {} {\bibfield  {journal} {\bibinfo
  {journal} {Phys. Rev. Fluids}\ }\textbf {\bibinfo {volume} {4}},\ \bibinfo
  {pages} {073101} (\bibinfo {year} {2019})}\BibitemShut {NoStop}%
\bibitem [{\citenamefont {Liu}\ \emph {et~al.}(2021)\citenamefont {Liu},
  \citenamefont {Zou}, \citenamefont {Tsang}, \citenamefont {Pak},\ and\
  \citenamefont {Young}}]{Liu2021}%
  \BibitemOpen
  \bibfield  {author} {\bibinfo {author} {\bibfnamefont {Y.}~\bibnamefont
  {Liu}}, \bibinfo {author} {\bibfnamefont {Z.}~\bibnamefont {Zou}}, \bibinfo
  {author} {\bibfnamefont {A.~C.~H.}\ \bibnamefont {Tsang}}, \bibinfo {author}
  {\bibfnamefont {O.~S.}\ \bibnamefont {Pak}},\ and\ \bibinfo {author}
  {\bibfnamefont {Y.-N.}\ \bibnamefont {Young}},\ }\bibfield  {title} {\enquote
  {\bibinfo {title} {{Mechanical rotation at low Reynolds number via
  reinforcement learning}},}\ }\href@noop {} {\bibfield  {journal} {\bibinfo
  {journal} {Physics of Fluids}\ }\textbf {\bibinfo {volume} {33}},\ \bibinfo
  {pages} {062007} (\bibinfo {year} {2021})}\BibitemShut {NoStop}%
\bibitem [{\citenamefont {Berdakin}, \citenamefont {Marconi},\ and\
  \citenamefont {Banchio}(2022)}]{Berdakin2022}%
  \BibitemOpen
  \bibfield  {author} {\bibinfo {author} {\bibfnamefont {I.}~\bibnamefont
  {Berdakin}}, \bibinfo {author} {\bibfnamefont {V.~I.}\ \bibnamefont
  {Marconi}},\ and\ \bibinfo {author} {\bibfnamefont {A.~J.}\ \bibnamefont
  {Banchio}},\ }\bibfield  {title} {\enquote {\bibinfo {title} {{Boosting
  micromachine studies with Stokesian dynamics}},}\ }\href@noop {} {\bibfield
  {journal} {\bibinfo  {journal} {Phys. Fluids}\ }\textbf {\bibinfo {volume}
  {34}},\ \bibinfo {pages} {037102} (\bibinfo {year} {2022})}\BibitemShut
  {NoStop}%
\bibitem [{\citenamefont {Leoni}\ \emph {et~al.}(2009)\citenamefont {Leoni},
  \citenamefont {Kotar}, \citenamefont {Bassetti}, \citenamefont {Cicuta},\
  and\ \citenamefont {Lagomarsino}}]{Leoni2009}%
  \BibitemOpen
  \bibfield  {author} {\bibinfo {author} {\bibfnamefont {M.}~\bibnamefont
  {Leoni}}, \bibinfo {author} {\bibfnamefont {J.}~\bibnamefont {Kotar}},
  \bibinfo {author} {\bibfnamefont {B.}~\bibnamefont {Bassetti}}, \bibinfo
  {author} {\bibfnamefont {P.}~\bibnamefont {Cicuta}},\ and\ \bibinfo {author}
  {\bibfnamefont {M.~C.}\ \bibnamefont {Lagomarsino}},\ }\bibfield  {title}
  {\enquote {\bibinfo {title} {{A basic swimmer at low Reynolds number}},}\
  }\href@noop {} {\bibfield  {journal} {\bibinfo  {journal} {Soft Matter}\
  }\textbf {\bibinfo {volume} {5}},\ \bibinfo {pages} {472--476} (\bibinfo
  {year} {2009})}\BibitemShut {NoStop}%
\bibitem [{\citenamefont {Grosjean}\ \emph {et~al.}(2016)\citenamefont
  {Grosjean}, \citenamefont {Hubert}, \citenamefont {Lagubeau},\ and\
  \citenamefont {Vandewalle}}]{Grosjean2016}%
  \BibitemOpen
  \bibfield  {author} {\bibinfo {author} {\bibfnamefont {G.}~\bibnamefont
  {Grosjean}}, \bibinfo {author} {\bibfnamefont {M.}~\bibnamefont {Hubert}},
  \bibinfo {author} {\bibfnamefont {G.}~\bibnamefont {Lagubeau}},\ and\
  \bibinfo {author} {\bibfnamefont {N.}~\bibnamefont {Vandewalle}},\ }\bibfield
   {title} {\enquote {\bibinfo {title} {{Realization of the Najafi-Golestanian
  microswimmer}},}\ }\href@noop {} {\bibfield  {journal} {\bibinfo  {journal}
  {Phys. Rev. E}\ }\textbf {\bibinfo {volume} {94}},\ \bibinfo {pages} {021101}
  (\bibinfo {year} {2016})}\BibitemShut {NoStop}%
\bibitem [{\citenamefont {Box}\ \emph {et~al.}(2017)\citenamefont {Box},
  \citenamefont {Han}, \citenamefont {Tipton},\ and\ \citenamefont
  {Mullin}}]{Box2017}%
  \BibitemOpen
  \bibfield  {author} {\bibinfo {author} {\bibfnamefont {F.}~\bibnamefont
  {Box}}, \bibinfo {author} {\bibfnamefont {E.}~\bibnamefont {Han}}, \bibinfo
  {author} {\bibfnamefont {C.~R.}\ \bibnamefont {Tipton}},\ and\ \bibinfo
  {author} {\bibfnamefont {T.}~\bibnamefont {Mullin}},\ }\bibfield  {title}
  {\enquote {\bibinfo {title} {{On the motion of linked spheres in a Stokes
  flow}},}\ }\href@noop {} {\bibfield  {journal} {\bibinfo  {journal} {Exp.
  Fluids}\ }\textbf {\bibinfo {volume} {58}},\ \bibinfo {pages} {29} (\bibinfo
  {year} {2017})}\BibitemShut {NoStop}%
\bibitem [{\citenamefont {Silverberg}\ \emph {et~al.}(2020)\citenamefont
  {Silverberg}, \citenamefont {Demir}, \citenamefont {Mishler}, \citenamefont
  {Hosoume}, \citenamefont {Trivedi}, \citenamefont {Tisch}, \citenamefont
  {Plascencia}, \citenamefont {Pak},\ and\ \citenamefont
  {Araci}}]{Silverberg_2020}%
  \BibitemOpen
  \bibfield  {author} {\bibinfo {author} {\bibfnamefont {O.}~\bibnamefont
  {Silverberg}}, \bibinfo {author} {\bibfnamefont {E.}~\bibnamefont {Demir}},
  \bibinfo {author} {\bibfnamefont {G.}~\bibnamefont {Mishler}}, \bibinfo
  {author} {\bibfnamefont {B.}~\bibnamefont {Hosoume}}, \bibinfo {author}
  {\bibfnamefont {N.}~\bibnamefont {Trivedi}}, \bibinfo {author} {\bibfnamefont
  {C.}~\bibnamefont {Tisch}}, \bibinfo {author} {\bibfnamefont
  {D.}~\bibnamefont {Plascencia}}, \bibinfo {author} {\bibfnamefont {O.~S.}\
  \bibnamefont {Pak}},\ and\ \bibinfo {author} {\bibfnamefont {I.~E.}\
  \bibnamefont {Araci}},\ }\bibfield  {title} {\enquote {\bibinfo {title}
  {{Realization of a push-me-pull-you swimmer at low Reynolds numbers}},}\
  }\href@noop {} {\bibfield  {journal} {\bibinfo  {journal} {Bioinspir.
  Biomim.}\ }\textbf {\bibinfo {volume} {15}},\ \bibinfo {pages} {064001}
  (\bibinfo {year} {2020})}\BibitemShut {NoStop}%
\bibitem [{\citenamefont {Curtis}\ and\ \citenamefont
  {Gaffney}(2013)}]{Curtis2013}%
  \BibitemOpen
  \bibfield  {author} {\bibinfo {author} {\bibfnamefont {M.~P.}\ \bibnamefont
  {Curtis}}\ and\ \bibinfo {author} {\bibfnamefont {E.~A.}\ \bibnamefont
  {Gaffney}},\ }\bibfield  {title} {\enquote {\bibinfo {title} {Three-sphere
  swimmer in a nonlinear viscoelastic medium},}\ }\href@noop {} {\bibfield
  {journal} {\bibinfo  {journal} {Phys. Rev. E}\ }\textbf {\bibinfo {volume}
  {87}},\ \bibinfo {pages} {043006} (\bibinfo {year} {2013})}\BibitemShut
  {NoStop}%
\bibitem [{\citenamefont {Yasuda}, \citenamefont {Hosaka},\ and\ \citenamefont
  {Komura}(2023)}]{yasuda2023generalized}%
  \BibitemOpen
  \bibfield  {author} {\bibinfo {author} {\bibfnamefont {K.}~\bibnamefont
  {Yasuda}}, \bibinfo {author} {\bibfnamefont {Y.}~\bibnamefont {Hosaka}},\
  and\ \bibinfo {author} {\bibfnamefont {S.}~\bibnamefont {Komura}},\
  }\bibfield  {title} {\enquote {\bibinfo {title} {Generalized three-sphere
  microswimmers},}\ }\href@noop {} {\bibfield  {journal} {\bibinfo  {journal}
  {arXiv preprint arXiv:2305.08411}\ } (\bibinfo {year} {2023})}\BibitemShut
  {NoStop}%
\bibitem [{\citenamefont {Pooley}, \citenamefont {Alexander},\ and\
  \citenamefont {Yeomans}(2007)}]{Pooley2007}%
  \BibitemOpen
  \bibfield  {author} {\bibinfo {author} {\bibfnamefont {C.~M.}\ \bibnamefont
  {Pooley}}, \bibinfo {author} {\bibfnamefont {G.~P.}\ \bibnamefont
  {Alexander}},\ and\ \bibinfo {author} {\bibfnamefont {J.~M.}\ \bibnamefont
  {Yeomans}},\ }\bibfield  {title} {\enquote {\bibinfo {title} {Hydrodynamic
  interaction between two swimmers at low reynolds number},}\ }\href@noop {}
  {\bibfield  {journal} {\bibinfo  {journal} {Phys. Rev. Lett.}\ }\textbf
  {\bibinfo {volume} {99}},\ \bibinfo {pages} {228103} (\bibinfo {year}
  {2007})}\BibitemShut {NoStop}%
\bibitem [{\citenamefont {Farzin}, \citenamefont {Ronasi},\ and\ \citenamefont
  {Najafi}(2012)}]{Farzin2012}%
  \BibitemOpen
  \bibfield  {author} {\bibinfo {author} {\bibfnamefont {M.}~\bibnamefont
  {Farzin}}, \bibinfo {author} {\bibfnamefont {K.}~\bibnamefont {Ronasi}},\
  and\ \bibinfo {author} {\bibfnamefont {A.}~\bibnamefont {Najafi}},\
  }\bibfield  {title} {\enquote {\bibinfo {title} {General aspects of
  hydrodynamic interactions between three-sphere low-reynolds-number
  swimmers},}\ }\href@noop {} {\bibfield  {journal} {\bibinfo  {journal} {Phys.
  Rev. E}\ }\textbf {\bibinfo {volume} {85}},\ \bibinfo {pages} {061914}
  (\bibinfo {year} {2012})}\BibitemShut {NoStop}%
\bibitem [{\citenamefont {Zargar}, \citenamefont {Najafi},\ and\ \citenamefont
  {Miri}(2009)}]{Zargar2009}%
  \BibitemOpen
  \bibfield  {author} {\bibinfo {author} {\bibfnamefont {R.}~\bibnamefont
  {Zargar}}, \bibinfo {author} {\bibfnamefont {A.}~\bibnamefont {Najafi}},\
  and\ \bibinfo {author} {\bibfnamefont {M.}~\bibnamefont {Miri}},\ }\bibfield
  {title} {\enquote {\bibinfo {title} {{Three-sphere low-Reynolds-number
  swimmer near a wall}},}\ }\href@noop {} {\bibfield  {journal} {\bibinfo
  {journal} {Phys. Rev. E}\ }\textbf {\bibinfo {volume} {80}},\ \bibinfo
  {pages} {026308} (\bibinfo {year} {2009})}\BibitemShut {NoStop}%
\bibitem [{\citenamefont {Najafi}, \citenamefont {Raad},\ and\ \citenamefont
  {Yousefi}(2013)}]{Najafi2013}%
  \BibitemOpen
  \bibfield  {author} {\bibinfo {author} {\bibfnamefont {A.}~\bibnamefont
  {Najafi}}, \bibinfo {author} {\bibfnamefont {S.~S.~H.}\ \bibnamefont
  {Raad}},\ and\ \bibinfo {author} {\bibfnamefont {R.}~\bibnamefont
  {Yousefi}},\ }\bibfield  {title} {\enquote {\bibinfo {title} {Self-propulsion
  in a low-reynolds-number fluid confined by two walls of a microchannel},}\
  }\href@noop {} {\bibfield  {journal} {\bibinfo  {journal} {Phys. Rev. E}\
  }\textbf {\bibinfo {volume} {88}},\ \bibinfo {pages} {045001} (\bibinfo
  {year} {2013})}\BibitemShut {NoStop}%
\bibitem [{\citenamefont {Daddi-Moussa-Ider}\ \emph
  {et~al.}(2018{\natexlab{a}})\citenamefont {Daddi-Moussa-Ider}, \citenamefont
  {Lisicki}, \citenamefont {Mathijssen}, \citenamefont {Hoell}, \citenamefont
  {Goh}, \citenamefont {Blawzdziewicz}, \citenamefont {Menzel},\ and\
  \citenamefont {L\"{o}wen}}]{Daddi-Moussa-Ider_2018}%
  \BibitemOpen
  \bibfield  {author} {\bibinfo {author} {\bibfnamefont {A.}~\bibnamefont
  {Daddi-Moussa-Ider}}, \bibinfo {author} {\bibfnamefont {M.}~\bibnamefont
  {Lisicki}}, \bibinfo {author} {\bibfnamefont {A.~J. T.~M.}\ \bibnamefont
  {Mathijssen}}, \bibinfo {author} {\bibfnamefont {C.}~\bibnamefont {Hoell}},
  \bibinfo {author} {\bibfnamefont {S.}~\bibnamefont {Goh}}, \bibinfo {author}
  {\bibfnamefont {J.}~\bibnamefont {Blawzdziewicz}}, \bibinfo {author}
  {\bibfnamefont {A.~M.}\ \bibnamefont {Menzel}},\ and\ \bibinfo {author}
  {\bibfnamefont {H.}~\bibnamefont {L\"{o}wen}},\ }\bibfield  {title} {\enquote
  {\bibinfo {title} {State diagram of a three-sphere microswimmer in a
  channel},}\ }\href@noop {} {\bibfield  {journal} {\bibinfo  {journal} {J.
  Phys.: Condens. Matter}\ }\textbf {\bibinfo {volume} {30}},\ \bibinfo {pages}
  {254004} (\bibinfo {year} {2018}{\natexlab{a}})}\BibitemShut {NoStop}%
\bibitem [{\citenamefont {Daddi-Moussa-Ider}\ \emph
  {et~al.}(2018{\natexlab{b}})\citenamefont {Daddi-Moussa-Ider}, \citenamefont
  {Lisicki}, \citenamefont {Hoell},\ and\ \citenamefont
  {Löwen}}]{DaddiMoussaIder2018_2}%
  \BibitemOpen
  \bibfield  {author} {\bibinfo {author} {\bibfnamefont {A.}~\bibnamefont
  {Daddi-Moussa-Ider}}, \bibinfo {author} {\bibfnamefont {M.}~\bibnamefont
  {Lisicki}}, \bibinfo {author} {\bibfnamefont {C.}~\bibnamefont {Hoell}},\
  and\ \bibinfo {author} {\bibfnamefont {H.}~\bibnamefont {Löwen}},\
  }\bibfield  {title} {\enquote {\bibinfo {title} {Swimming trajectories of a
  three-sphere microswimmer near a wall},}\ }\href@noop {} {\bibfield
  {journal} {\bibinfo  {journal} {J. Chem. Phys.}\ }\textbf {\bibinfo {volume}
  {148}},\ \bibinfo {pages} {134904} (\bibinfo {year}
  {2018}{\natexlab{b}})}\BibitemShut {NoStop}%
\bibitem [{\citenamefont {Daddi-Moussa-Ider}, \citenamefont {Lisicki},\ and\
  \citenamefont {Mathijssen}(2020)}]{daddi2020tuning}%
  \BibitemOpen
  \bibfield  {author} {\bibinfo {author} {\bibfnamefont {A.}~\bibnamefont
  {Daddi-Moussa-Ider}}, \bibinfo {author} {\bibfnamefont {M.}~\bibnamefont
  {Lisicki}},\ and\ \bibinfo {author} {\bibfnamefont {A.~J.}\ \bibnamefont
  {Mathijssen}},\ }\bibfield  {title} {\enquote {\bibinfo {title} {Tuning the
  upstream swimming of microrobots by shape and cargo size},}\ }\href@noop {}
  {\bibfield  {journal} {\bibinfo  {journal} {Phys. Rev. Applied}\ }\textbf
  {\bibinfo {volume} {14}},\ \bibinfo {pages} {024071} (\bibinfo {year}
  {2020})}\BibitemShut {NoStop}%
\bibitem [{\citenamefont {Tsang}\ \emph
  {et~al.}(2020{\natexlab{b}})\citenamefont {Tsang}, \citenamefont {Tong},
  \citenamefont {Nallan},\ and\ \citenamefont {Pak}}]{AlanChengHouTsang2020}%
  \BibitemOpen
  \bibfield  {author} {\bibinfo {author} {\bibfnamefont {A.~C.~H.}\
  \bibnamefont {Tsang}}, \bibinfo {author} {\bibfnamefont {P.~W.}\ \bibnamefont
  {Tong}}, \bibinfo {author} {\bibfnamefont {S.}~\bibnamefont {Nallan}},\ and\
  \bibinfo {author} {\bibfnamefont {O.~S.}\ \bibnamefont {Pak}},\ }\bibfield
  {title} {\enquote {\bibinfo {title} {{Self-learning how to swim at low
  Reynolds number}},}\ }\href@noop {} {\bibfield  {journal} {\bibinfo
  {journal} {Phys. Rev. Fluids}\ }\textbf {\bibinfo {volume} {5}},\ \bibinfo
  {pages} {074101} (\bibinfo {year} {2020}{\natexlab{b}})}\BibitemShut
  {NoStop}%
\bibitem [{\citenamefont {Hartl}\ \emph {et~al.}(2021)\citenamefont {Hartl},
  \citenamefont {Hübl}, \citenamefont {Kahl},\ and\ \citenamefont
  {Zöttl}}]{Hartl2021}%
  \BibitemOpen
  \bibfield  {author} {\bibinfo {author} {\bibfnamefont {B.}~\bibnamefont
  {Hartl}}, \bibinfo {author} {\bibfnamefont {M.}~\bibnamefont {Hübl}},
  \bibinfo {author} {\bibfnamefont {G.}~\bibnamefont {Kahl}},\ and\ \bibinfo
  {author} {\bibfnamefont {A.}~\bibnamefont {Zöttl}},\ }\bibfield  {title}
  {\enquote {\bibinfo {title} {Microswimmers learning chemotaxis with genetic
  algorithms},}\ }\href@noop {} {\bibfield  {journal} {\bibinfo  {journal}
  {Proc. Natl. Acad. Sci. U.S.A.}\ }\textbf {\bibinfo {volume} {118}},\
  \bibinfo {pages} {e2019683118} (\bibinfo {year} {2021})}\BibitemShut
  {NoStop}%
\bibitem [{\citenamefont {Zou}\ \emph {et~al.}(2022)\citenamefont {Zou},
  \citenamefont {Liu}, \citenamefont {Young}, \citenamefont {Pak},\ and\
  \citenamefont {Tsang}}]{Zou2022}%
  \BibitemOpen
  \bibfield  {author} {\bibinfo {author} {\bibfnamefont {Z.}~\bibnamefont
  {Zou}}, \bibinfo {author} {\bibfnamefont {Y.}~\bibnamefont {Liu}}, \bibinfo
  {author} {\bibfnamefont {Y.~N.}\ \bibnamefont {Young}}, \bibinfo {author}
  {\bibfnamefont {O.~S.}\ \bibnamefont {Pak}},\ and\ \bibinfo {author}
  {\bibfnamefont {A.~C.~H.}\ \bibnamefont {Tsang}},\ }\bibfield  {title}
  {\enquote {\bibinfo {title} {Gait switching and targeted navigation of
  microswimmers via deep reinforcement learning},}\ }\href@noop {} {\bibfield
  {journal} {\bibinfo  {journal} {Commun. Phys.}\ }\textbf {\bibinfo {volume}
  {5}},\ \bibinfo {pages} {158} (\bibinfo {year} {2022})}\BibitemShut {NoStop}%
\bibitem [{\citenamefont {Paz}\ \emph {et~al.}(2023)\citenamefont {Paz},
  \citenamefont {Ausas}, \citenamefont {Carbajal},\ and\ \citenamefont
  {Buscaglia}}]{Paz2023}%
  \BibitemOpen
  \bibfield  {author} {\bibinfo {author} {\bibfnamefont {S.}~\bibnamefont
  {Paz}}, \bibinfo {author} {\bibfnamefont {R.~F.}\ \bibnamefont {Ausas}},
  \bibinfo {author} {\bibfnamefont {J.~P.}\ \bibnamefont {Carbajal}},\ and\
  \bibinfo {author} {\bibfnamefont {G.~C.}\ \bibnamefont {Buscaglia}},\
  }\bibfield  {title} {\enquote {\bibinfo {title} {Chemoreception and
  chemotaxis of a three-sphere swimmer},}\ }\href@noop {} {\bibfield  {journal}
  {\bibinfo  {journal} {Commun. Nonlinear Sci. Numer. Simul.}\ }\textbf
  {\bibinfo {volume} {117}},\ \bibinfo {pages} {106909} (\bibinfo {year}
  {2023})}\BibitemShut {NoStop}%
\bibitem [{\citenamefont {Liu}\ \emph {et~al.}(2023)\citenamefont {Liu},
  \citenamefont {Zou}, \citenamefont {Pak},\ and\ \citenamefont
  {Tsang}}]{Liu2023}%
  \BibitemOpen
  \bibfield  {author} {\bibinfo {author} {\bibfnamefont {Y.}~\bibnamefont
  {Liu}}, \bibinfo {author} {\bibfnamefont {Z.}~\bibnamefont {Zou}}, \bibinfo
  {author} {\bibfnamefont {O.~S.}\ \bibnamefont {Pak}},\ and\ \bibinfo {author}
  {\bibfnamefont {A.~C.~H.}\ \bibnamefont {Tsang}},\ }\bibfield  {title}
  {\enquote {\bibinfo {title} {Learning to cooperate for low-reynolds-number
  swimming: a model problem for gait coordination},}\ }\href@noop {} {\bibfield
   {journal} {\bibinfo  {journal} {Sci. Rep.}\ }\textbf {\bibinfo {volume}
  {13}},\ \bibinfo {pages} {9397} (\bibinfo {year} {2023})}\BibitemShut
  {NoStop}%
\bibitem [{\citenamefont {Katz}(1974)}]{katz_1974}%
  \BibitemOpen
  \bibfield  {author} {\bibinfo {author} {\bibfnamefont {D.~F.}\ \bibnamefont
  {Katz}},\ }\bibfield  {title} {\enquote {\bibinfo {title} {On the propulsion
  of micro-organisms near solid boundaries},}\ }\href@noop {} {\bibfield
  {journal} {\bibinfo  {journal} {J. Fluid Mech.}\ }\textbf {\bibinfo {volume}
  {64}},\ \bibinfo {pages} {33–49} (\bibinfo {year} {1974})}\BibitemShut
  {NoStop}%
\bibitem [{\citenamefont {Fauci}\ and\ \citenamefont
  {McDonald}(1995)}]{Fauci1995}%
  \BibitemOpen
  \bibfield  {author} {\bibinfo {author} {\bibfnamefont {L.~J.}\ \bibnamefont
  {Fauci}}\ and\ \bibinfo {author} {\bibfnamefont {A.}~\bibnamefont
  {McDonald}},\ }\bibfield  {title} {\enquote {\bibinfo {title} {Sperm motility
  in the presence of boundaries},}\ }\href@noop {} {\bibfield  {journal}
  {\bibinfo  {journal} {Bull. Math. Biol.}\ }\textbf {\bibinfo {volume} {57}},\
  \bibinfo {pages} {679--699} (\bibinfo {year} {1995})}\BibitemShut {NoStop}%
\bibitem [{\citenamefont {Hernandez-Ortiz}, \citenamefont {Stoltz},\ and\
  \citenamefont {Graham}(2005)}]{Ortiz2005}%
  \BibitemOpen
  \bibfield  {author} {\bibinfo {author} {\bibfnamefont {J.~P.}\ \bibnamefont
  {Hernandez-Ortiz}}, \bibinfo {author} {\bibfnamefont {C.~G.}\ \bibnamefont
  {Stoltz}},\ and\ \bibinfo {author} {\bibfnamefont {M.~D.}\ \bibnamefont
  {Graham}},\ }\bibfield  {title} {\enquote {\bibinfo {title} {Transport and
  collective dynamics in suspensions of confined swimming particles},}\
  }\href@noop {} {\bibfield  {journal} {\bibinfo  {journal} {Phys. Rev. Lett.}\
  }\textbf {\bibinfo {volume} {95}},\ \bibinfo {pages} {204501} (\bibinfo
  {year} {2005})}\BibitemShut {NoStop}%
\bibitem [{\citenamefont {Lauga}\ \emph {et~al.}(2006)\citenamefont {Lauga},
  \citenamefont {DiLuzio}, \citenamefont {Whitesides},\ and\ \citenamefont
  {Stone}}]{Lauga2006}%
  \BibitemOpen
  \bibfield  {author} {\bibinfo {author} {\bibfnamefont {E.}~\bibnamefont
  {Lauga}}, \bibinfo {author} {\bibfnamefont {W.~R.}\ \bibnamefont {DiLuzio}},
  \bibinfo {author} {\bibfnamefont {G.~M.}\ \bibnamefont {Whitesides}},\ and\
  \bibinfo {author} {\bibfnamefont {H.~A.}\ \bibnamefont {Stone}},\ }\bibfield
  {title} {\enquote {\bibinfo {title} {Swimming in circles: Motion of bacteria
  near solid boundaries},}\ }\href@noop {} {\bibfield  {journal} {\bibinfo
  {journal} {Biophys. J.}\ }\textbf {\bibinfo {volume} {90}},\ \bibinfo {pages}
  {400--412} (\bibinfo {year} {2006})}\BibitemShut {NoStop}%
\bibitem [{\citenamefont {Smith}\ \emph {et~al.}(2009)\citenamefont {Smith},
  \citenamefont {Gaffney}, \citenamefont {Blake},\ and\ \citenamefont
  {Kirkman-Brown}}]{smith2009}%
  \BibitemOpen
  \bibfield  {author} {\bibinfo {author} {\bibfnamefont {D.}~\bibnamefont
  {Smith}}, \bibinfo {author} {\bibfnamefont {E.}~\bibnamefont {Gaffney}},
  \bibinfo {author} {\bibfnamefont {J.}~\bibnamefont {Blake}},\ and\ \bibinfo
  {author} {\bibfnamefont {J.}~\bibnamefont {Kirkman-Brown}},\ }\bibfield
  {title} {\enquote {\bibinfo {title} {Human sperm accumulation near surfaces:
  a simulation study},}\ }\href@noop {} {\bibfield  {journal} {\bibinfo
  {journal} {J. Fluid Mech.}\ }\textbf {\bibinfo {volume} {621}},\ \bibinfo
  {pages} {289--320} (\bibinfo {year} {2009})}\BibitemShut {NoStop}%
\bibitem [{\citenamefont {Giacch\'e}, \citenamefont {Ishikawa},\ and\
  \citenamefont {Yamaguchi}(2010)}]{Davide2010}%
  \BibitemOpen
  \bibfield  {author} {\bibinfo {author} {\bibfnamefont {D.}~\bibnamefont
  {Giacch\'e}}, \bibinfo {author} {\bibfnamefont {T.}~\bibnamefont
  {Ishikawa}},\ and\ \bibinfo {author} {\bibfnamefont {T.}~\bibnamefont
  {Yamaguchi}},\ }\bibfield  {title} {\enquote {\bibinfo {title} {Hydrodynamic
  entrapment of bacteria swimming near a solid surface},}\ }\href@noop {}
  {\bibfield  {journal} {\bibinfo  {journal} {Phys. Rev. E}\ }\textbf {\bibinfo
  {volume} {82}},\ \bibinfo {pages} {056309} (\bibinfo {year}
  {2010})}\BibitemShut {NoStop}%
\bibitem [{\citenamefont {Shum}, \citenamefont {Gaffney},\ and\ \citenamefont
  {Smith}(2010)}]{Shum2010}%
  \BibitemOpen
  \bibfield  {author} {\bibinfo {author} {\bibfnamefont {H.}~\bibnamefont
  {Shum}}, \bibinfo {author} {\bibfnamefont {E.~A.}\ \bibnamefont {Gaffney}},\
  and\ \bibinfo {author} {\bibfnamefont {D.~J.}\ \bibnamefont {Smith}},\
  }\bibfield  {title} {\enquote {\bibinfo {title} {Modelling bacterial
  behaviour close to a no-slip plane boundary: the influence of bacterial
  geometry},}\ }\href@noop {} {\bibfield  {journal} {\bibinfo  {journal} {Proc.
  R. Soc. A}\ }\textbf {\bibinfo {volume} {466}},\ \bibinfo {pages}
  {1725--1748} (\bibinfo {year} {2010})}\BibitemShut {NoStop}%
\bibitem [{\citenamefont {Crowdy}\ and\ \citenamefont {Or}(2010)}]{Crowdy2010}%
  \BibitemOpen
  \bibfield  {author} {\bibinfo {author} {\bibfnamefont {D.~G.}\ \bibnamefont
  {Crowdy}}\ and\ \bibinfo {author} {\bibfnamefont {Y.}~\bibnamefont {Or}},\
  }\bibfield  {title} {\enquote {\bibinfo {title} {{Two-dimensional point
  singularity model of a low-Reynolds-number swimmer near a wall}},}\
  }\href@noop {} {\bibfield  {journal} {\bibinfo  {journal} {Phys. Rev. E}\
  }\textbf {\bibinfo {volume} {81}},\ \bibinfo {pages} {036313} (\bibinfo
  {year} {2010})}\BibitemShut {NoStop}%
\bibitem [{\citenamefont {Z\"ottl}\ and\ \citenamefont
  {Stark}(2012)}]{Andreas2012}%
  \BibitemOpen
  \bibfield  {author} {\bibinfo {author} {\bibfnamefont {A.}~\bibnamefont
  {Z\"ottl}}\ and\ \bibinfo {author} {\bibfnamefont {H.}~\bibnamefont
  {Stark}},\ }\bibfield  {title} {\enquote {\bibinfo {title} {Nonlinear
  dynamics of a microswimmer in poiseuille flow},}\ }\href@noop {} {\bibfield
  {journal} {\bibinfo  {journal} {Phys. Rev. Lett.}\ }\textbf {\bibinfo
  {volume} {108}},\ \bibinfo {pages} {218104} (\bibinfo {year}
  {2012})}\BibitemShut {NoStop}%
\bibitem [{\citenamefont {Spagnolie}\ and\ \citenamefont
  {Lauga}(2012)}]{spagnolie_lauga_2012}%
  \BibitemOpen
  \bibfield  {author} {\bibinfo {author} {\bibfnamefont {S.~E.}\ \bibnamefont
  {Spagnolie}}\ and\ \bibinfo {author} {\bibfnamefont {E.}~\bibnamefont
  {Lauga}},\ }\bibfield  {title} {\enquote {\bibinfo {title} {Hydrodynamics of
  self-propulsion near a boundary: predictions and accuracy of far-field
  approximations},}\ }\href@noop {} {\bibfield  {journal} {\bibinfo  {journal}
  {J. Fluid Mech.}\ }\textbf {\bibinfo {volume} {700}},\ \bibinfo {pages}
  {105–147} (\bibinfo {year} {2012})}\BibitemShut {NoStop}%
\bibitem [{\citenamefont {Li}\ and\ \citenamefont {Ardekani}(2014)}]{Li2014}%
  \BibitemOpen
  \bibfield  {author} {\bibinfo {author} {\bibfnamefont {G.-J.}\ \bibnamefont
  {Li}}\ and\ \bibinfo {author} {\bibfnamefont {A.~M.}\ \bibnamefont
  {Ardekani}},\ }\bibfield  {title} {\enquote {\bibinfo {title} {Hydrodynamic
  interaction of microswimmers near a wall},}\ }\href@noop {} {\bibfield
  {journal} {\bibinfo  {journal} {Phys. Rev. E}\ }\textbf {\bibinfo {volume}
  {90}},\ \bibinfo {pages} {013010} (\bibinfo {year} {2014})}\BibitemShut
  {NoStop}%
\bibitem [{\citenamefont {Bayati}\ \emph {et~al.}(2019)\citenamefont {Bayati},
  \citenamefont {Popescu}, \citenamefont {Uspal}, \citenamefont {Dietrich},\
  and\ \citenamefont {Najafi}}]{Bayati2019}%
  \BibitemOpen
  \bibfield  {author} {\bibinfo {author} {\bibfnamefont {P.}~\bibnamefont
  {Bayati}}, \bibinfo {author} {\bibfnamefont {M.~N.}\ \bibnamefont {Popescu}},
  \bibinfo {author} {\bibfnamefont {W.~E.}\ \bibnamefont {Uspal}}, \bibinfo
  {author} {\bibfnamefont {S.}~\bibnamefont {Dietrich}},\ and\ \bibinfo
  {author} {\bibfnamefont {A.}~\bibnamefont {Najafi}},\ }\bibfield  {title}
  {\enquote {\bibinfo {title} {Dynamics near planar walls for various model
  self-phoretic particles},}\ }\href@noop {} {\bibfield  {journal} {\bibinfo
  {journal} {Soft Matter}\ }\textbf {\bibinfo {volume} {15}},\ \bibinfo {pages}
  {5644--5672} (\bibinfo {year} {2019})}\BibitemShut {NoStop}%
\bibitem [{\citenamefont {Farutin}\ \emph {et~al.}(2019)\citenamefont
  {Farutin}, \citenamefont {Wu}, \citenamefont {Hu}, \citenamefont {Rafaï},
  \citenamefont {Peyla}, \citenamefont {Lai},\ and\ \citenamefont
  {Misbah}}]{Farutin2019}%
  \BibitemOpen
  \bibfield  {author} {\bibinfo {author} {\bibfnamefont {A.}~\bibnamefont
  {Farutin}}, \bibinfo {author} {\bibfnamefont {H.}~\bibnamefont {Wu}},
  \bibinfo {author} {\bibfnamefont {W.~F.}\ \bibnamefont {Hu}}, \bibinfo
  {author} {\bibfnamefont {S.}~\bibnamefont {Rafaï}}, \bibinfo {author}
  {\bibfnamefont {P.}~\bibnamefont {Peyla}}, \bibinfo {author} {\bibfnamefont
  {M.~C.}\ \bibnamefont {Lai}},\ and\ \bibinfo {author} {\bibfnamefont
  {C.}~\bibnamefont {Misbah}},\ }\bibfield  {title} {\enquote {\bibinfo {title}
  {Analytical study for swimmers in a channel},}\ }\href@noop {} {\bibfield
  {journal} {\bibinfo  {journal} {J. Fluid Mech.}\ }\textbf {\bibinfo {volume}
  {881}},\ \bibinfo {pages} {365--383} (\bibinfo {year} {2019})}\BibitemShut
  {NoStop}%
\bibitem [{\citenamefont {Felderhof}(2010)}]{Felderhof2010}%
  \BibitemOpen
  \bibfield  {author} {\bibinfo {author} {\bibfnamefont {B.~U.}\ \bibnamefont
  {Felderhof}},\ }\bibfield  {title} {\enquote {\bibinfo {title} {{Swimming at
  low Reynolds number of a cylindrical body in a circular tube}},}\ }\href@noop
  {} {\bibfield  {journal} {\bibinfo  {journal} {Phys. Fluids}\ }\textbf
  {\bibinfo {volume} {22}},\ \bibinfo {pages} {113604} (\bibinfo {year}
  {2010})}\BibitemShut {NoStop}%
\bibitem [{\citenamefont {Jana}, \citenamefont {Um},\ and\ \citenamefont
  {Jung}(2012)}]{Jana2012}%
  \BibitemOpen
  \bibfield  {author} {\bibinfo {author} {\bibfnamefont {S.}~\bibnamefont
  {Jana}}, \bibinfo {author} {\bibfnamefont {S.~H.}\ \bibnamefont {Um}},\ and\
  \bibinfo {author} {\bibfnamefont {S.}~\bibnamefont {Jung}},\ }\bibfield
  {title} {\enquote {\bibinfo {title} {{Paramecium swimming in capillary
  tube}},}\ }\href@noop {} {\bibfield  {journal} {\bibinfo  {journal} {Phys.
  Fluids}\ }\textbf {\bibinfo {volume} {24}},\ \bibinfo {pages} {041901}
  (\bibinfo {year} {2012})}\BibitemShut {NoStop}%
\bibitem [{\citenamefont {Zhu}, \citenamefont {Lauga},\ and\ \citenamefont
  {Brandt}(2013)}]{Zhu2013}%
  \BibitemOpen
  \bibfield  {author} {\bibinfo {author} {\bibfnamefont {L.}~\bibnamefont
  {Zhu}}, \bibinfo {author} {\bibfnamefont {E.}~\bibnamefont {Lauga}},\ and\
  \bibinfo {author} {\bibfnamefont {L.}~\bibnamefont {Brandt}},\ }\bibfield
  {title} {\enquote {\bibinfo {title} {{Low-Reynolds-number swimming in a
  capillary tube}},}\ }\href@noop {} {\bibfield  {journal} {\bibinfo  {journal}
  {J. Fluid Mech.}\ }\textbf {\bibinfo {volume} {726}},\ \bibinfo {pages}
  {285--311} (\bibinfo {year} {2013})}\BibitemShut {NoStop}%
\bibitem [{\citenamefont {Ledesma-Aguilar}\ and\ \citenamefont
  {Yeomans}(2013)}]{LedesmaAguilar2013}%
  \BibitemOpen
  \bibfield  {author} {\bibinfo {author} {\bibfnamefont {R.}~\bibnamefont
  {Ledesma-Aguilar}}\ and\ \bibinfo {author} {\bibfnamefont {J.~M.}\
  \bibnamefont {Yeomans}},\ }\bibfield  {title} {\enquote {\bibinfo {title}
  {Enhanced motility of a microswimmer in rigid and elastic confinement},}\
  }\href@noop {} {\bibfield  {journal} {\bibinfo  {journal} {Phys. Rev. Lett.}\
  }\textbf {\bibinfo {volume} {111}},\ \bibinfo {pages} {138101} (\bibinfo
  {year} {2013})}\BibitemShut {NoStop}%
\bibitem [{\citenamefont {Liu}, \citenamefont {Breuer},\ and\ \citenamefont
  {Powers}(2014)}]{Liu2014}%
  \BibitemOpen
  \bibfield  {author} {\bibinfo {author} {\bibfnamefont {B.}~\bibnamefont
  {Liu}}, \bibinfo {author} {\bibfnamefont {K.~S.}\ \bibnamefont {Breuer}},\
  and\ \bibinfo {author} {\bibfnamefont {T.~R.}\ \bibnamefont {Powers}},\
  }\bibfield  {title} {\enquote {\bibinfo {title} {{Propulsion by a helical
  flagellum in a capillary tube}},}\ }\href@noop {} {\bibfield  {journal}
  {\bibinfo  {journal} {Phys. Fluids}\ }\textbf {\bibinfo {volume} {26}},\
  \bibinfo {pages} {011701} (\bibinfo {year} {2014})}\BibitemShut {NoStop}%
\bibitem [{\citenamefont {Caldag}\ and\ \citenamefont
  {Yesilyurt}(2019)}]{Caldag2019}%
  \BibitemOpen
  \bibfield  {author} {\bibinfo {author} {\bibfnamefont {H.~O.}\ \bibnamefont
  {Caldag}}\ and\ \bibinfo {author} {\bibfnamefont {S.}~\bibnamefont
  {Yesilyurt}},\ }\bibfield  {title} {\enquote {\bibinfo {title} {{Trajectories
  of magnetically-actuated helical swimmers in cylindrical channels at low
  Reynolds numbers}},}\ }\href@noop {} {\bibfield  {journal} {\bibinfo
  {journal} {J. Fluids Struct.}\ }\textbf {\bibinfo {volume} {90}},\ \bibinfo
  {pages} {164--176} (\bibinfo {year} {2019})}\BibitemShut {NoStop}%
\bibitem [{\citenamefont {Ouyang}\ \emph {et~al.}(2022)\citenamefont {Ouyang},
  \citenamefont {Lin}, \citenamefont {Yu}, \citenamefont {Lin},\ and\
  \citenamefont {Phan-Thien}}]{ouyang_lin_yu_lin_phan-thien_2022}%
  \BibitemOpen
  \bibfield  {author} {\bibinfo {author} {\bibfnamefont {Z.}~\bibnamefont
  {Ouyang}}, \bibinfo {author} {\bibfnamefont {Z.}~\bibnamefont {Lin}},
  \bibinfo {author} {\bibfnamefont {Z.}~\bibnamefont {Yu}}, \bibinfo {author}
  {\bibfnamefont {J.}~\bibnamefont {Lin}},\ and\ \bibinfo {author}
  {\bibfnamefont {N.}~\bibnamefont {Phan-Thien}},\ }\bibfield  {title}
  {\enquote {\bibinfo {title} {Hydrodynamics of an inertial squirmer and
  squirmer dumbbell in a tube},}\ }\href@noop {} {\bibfield  {journal}
  {\bibinfo  {journal} {J. Fluid Mech.}\ }\textbf {\bibinfo {volume} {939}},\
  \bibinfo {pages} {A32} (\bibinfo {year} {2022})}\BibitemShut {NoStop}%
\bibitem [{\citenamefont {Golestanian}\ and\ \citenamefont
  {Ajdari}(2008)}]{GolestanianAjdari2008}%
  \BibitemOpen
  \bibfield  {author} {\bibinfo {author} {\bibfnamefont {R.}~\bibnamefont
  {Golestanian}}\ and\ \bibinfo {author} {\bibfnamefont {A.}~\bibnamefont
  {Ajdari}},\ }\bibfield  {title} {\enquote {\bibinfo {title} {{Analytic
  results for the three-sphere swimmer at low Reynolds number}},}\ }\href@noop
  {} {\bibfield  {journal} {\bibinfo  {journal} {Phys. Rev. E}\ }\textbf
  {\bibinfo {volume} {77}},\ \bibinfo {pages} {036308} (\bibinfo {year}
  {2008})}\BibitemShut {NoStop}%
\bibitem [{\citenamefont {Bohlin}(1960)}]{bohlin1960drag}%
  \BibitemOpen
  \bibfield  {author} {\bibinfo {author} {\bibfnamefont {T.}~\bibnamefont
  {Bohlin}},\ }\bibfield  {title} {\enquote {\bibinfo {title} {On the drag on a
  rigid sphere moving in a viscous liquid inside a cylindrical tube},}\
  }\href@noop {} {\bibfield  {journal} {\bibinfo  {journal} {Trans. Roy. Inst.
  Technol. Stockholm}\ }\textbf {\bibinfo {volume} {155}},\ \bibinfo {pages}
  {64} (\bibinfo {year} {1960})}\BibitemShut {NoStop}%
\bibitem [{\citenamefont {Daddi-Moussa-Ider}, \citenamefont {Lisicki},\ and\
  \citenamefont {Gekle}(2017)}]{daddi2017hydrodynamic}%
  \BibitemOpen
  \bibfield  {author} {\bibinfo {author} {\bibfnamefont {A.}~\bibnamefont
  {Daddi-Moussa-Ider}}, \bibinfo {author} {\bibfnamefont {M.}~\bibnamefont
  {Lisicki}},\ and\ \bibinfo {author} {\bibfnamefont {S.}~\bibnamefont
  {Gekle}},\ }\bibfield  {title} {\enquote {\bibinfo {title} {Hydrodynamic
  mobility of a sphere moving on the centerline of an elastic tube},}\
  }\href@noop {} {\bibfield  {journal} {\bibinfo  {journal} {Phys. Fluids}\
  }\textbf {\bibinfo {volume} {29}},\ \bibinfo {pages} {111901} (\bibinfo
  {year} {2017})}\BibitemShut {NoStop}%
\bibitem [{\citenamefont {Daddi Moussa~Ider}(2017)}]{daddi2017diffusion}%
  \BibitemOpen
  \bibfield  {author} {\bibinfo {author} {\bibfnamefont {A.}~\bibnamefont
  {Daddi Moussa~Ider}},\ }\emph {\bibinfo {title} {Diffusion of nanoparticles
  nearby elastic cell membranes: A theoretical study}},\ \href@noop {} {Ph.D.
  thesis},\ \bibinfo  {school} {Fakultät für Mathematik, Physik und
  Informatik, Universität Bayreuth, Germany} (\bibinfo {year}
  {2017})\BibitemShut {NoStop}%
\bibitem [{\citenamefont {Cui}, \citenamefont {Diamant},\ and\ \citenamefont
  {Lin}(2002)}]{cui2002screened}%
  \BibitemOpen
  \bibfield  {author} {\bibinfo {author} {\bibfnamefont {B.}~\bibnamefont
  {Cui}}, \bibinfo {author} {\bibfnamefont {H.}~\bibnamefont {Diamant}},\ and\
  \bibinfo {author} {\bibfnamefont {B.}~\bibnamefont {Lin}},\ }\bibfield
  {title} {\enquote {\bibinfo {title} {Screened hydrodynamic interaction in a
  narrow channel},}\ }\href@noop {} {\bibfield  {journal} {\bibinfo  {journal}
  {Phys. Rev. Lett.}\ }\textbf {\bibinfo {volume} {89}},\ \bibinfo {pages}
  {188302} (\bibinfo {year} {2002})}\BibitemShut {NoStop}%
\bibitem [{\citenamefont {Liron}\ and\ \citenamefont
  {Shahar}(1978)}]{liron_shahar_1978}%
  \BibitemOpen
  \bibfield  {author} {\bibinfo {author} {\bibfnamefont {N.}~\bibnamefont
  {Liron}}\ and\ \bibinfo {author} {\bibfnamefont {R.}~\bibnamefont {Shahar}},\
  }\bibfield  {title} {\enquote {\bibinfo {title} {{Stokes flow due to a
  Stokeslet in a pipe}},}\ }\href@noop {} {\bibfield  {journal} {\bibinfo
  {journal} {J. Fluid Mech.}\ }\textbf {\bibinfo {volume} {86}},\ \bibinfo
  {pages} {727–744} (\bibinfo {year} {1978})}\BibitemShut {NoStop}%
\bibitem [{\citenamefont {Leshansky}(2009)}]{Leshansky2009}%
  \BibitemOpen
  \bibfield  {author} {\bibinfo {author} {\bibfnamefont {A.~M.}\ \bibnamefont
  {Leshansky}},\ }\bibfield  {title} {\enquote {\bibinfo {title} {{Enhanced
  low-Reynolds-number propulsion in heterogeneous viscous environments}},}\
  }\href@noop {} {\bibfield  {journal} {\bibinfo  {journal} {Phys. Rev. E}\
  }\textbf {\bibinfo {volume} {80}},\ \bibinfo {pages} {051911} (\bibinfo
  {year} {2009})}\BibitemShut {NoStop}%
\bibitem [{\citenamefont {Fu}, \citenamefont {Shenoy},\ and\ \citenamefont
  {Powers}(2010)}]{Fu_2010}%
  \BibitemOpen
  \bibfield  {author} {\bibinfo {author} {\bibfnamefont {H.~C.}\ \bibnamefont
  {Fu}}, \bibinfo {author} {\bibfnamefont {V.~B.}\ \bibnamefont {Shenoy}},\
  and\ \bibinfo {author} {\bibfnamefont {T.~R.}\ \bibnamefont {Powers}},\
  }\bibfield  {title} {\enquote {\bibinfo {title} {{Low-Reynolds-number
  swimming in gels}},}\ }\href@noop {} {\bibfield  {journal} {\bibinfo
  {journal} {EPL}\ }\textbf {\bibinfo {volume} {91}},\ \bibinfo {pages} {24002}
  (\bibinfo {year} {2010})}\BibitemShut {NoStop}%
\bibitem [{\citenamefont {Ho}, \citenamefont {Olson},\ and\ \citenamefont
  {Leiderman}(2016)}]{Nguyenho2016}%
  \BibitemOpen
  \bibfield  {author} {\bibinfo {author} {\bibfnamefont {N.}~\bibnamefont
  {Ho}}, \bibinfo {author} {\bibfnamefont {S.~D.}\ \bibnamefont {Olson}},\ and\
  \bibinfo {author} {\bibfnamefont {K.}~\bibnamefont {Leiderman}},\ }\bibfield
  {title} {\enquote {\bibinfo {title} {Swimming speeds of filaments in viscous
  fluids with resistance},}\ }\href@noop {} {\bibfield  {journal} {\bibinfo
  {journal} {Phys. Rev. E}\ }\textbf {\bibinfo {volume} {93}},\ \bibinfo
  {pages} {043108} (\bibinfo {year} {2016})}\BibitemShut {NoStop}%
\bibitem [{\citenamefont {Leiderman}\ and\ \citenamefont
  {Olson}(2016)}]{Leiderman2016}%
  \BibitemOpen
  \bibfield  {author} {\bibinfo {author} {\bibfnamefont {K.}~\bibnamefont
  {Leiderman}}\ and\ \bibinfo {author} {\bibfnamefont {S.~D.}\ \bibnamefont
  {Olson}},\ }\bibfield  {title} {\enquote {\bibinfo {title} {{Swimming in a
  two-dimensional Brinkman fluid: Computational modeling and regularized
  solutions}},}\ }\href@noop {} {\bibfield  {journal} {\bibinfo  {journal}
  {Phys. Fluids}\ }\textbf {\bibinfo {volume} {28}},\ \bibinfo {pages} {021902}
  (\bibinfo {year} {2016})}\BibitemShut {NoStop}%
\bibitem [{\citenamefont {Daddi-Moussa-Ider}\ and\ \citenamefont
  {Menzel}(2018)}]{PhysRevFluids.3.094102}%
  \BibitemOpen
  \bibfield  {author} {\bibinfo {author} {\bibfnamefont {A.}~\bibnamefont
  {Daddi-Moussa-Ider}}\ and\ \bibinfo {author} {\bibfnamefont {A.~M.}\
  \bibnamefont {Menzel}},\ }\bibfield  {title} {\enquote {\bibinfo {title}
  {Dynamics of a simple model microswimmer in an anisotropic fluid:
  Implications for alignment behavior and active transport in a nematic liquid
  crystal},}\ }\href@noop {} {\bibfield  {journal} {\bibinfo  {journal} {Phys.
  Rev. Fluids}\ }\textbf {\bibinfo {volume} {3}},\ \bibinfo {pages} {094102}
  (\bibinfo {year} {2018})}\BibitemShut {NoStop}%
\bibitem [{\citenamefont {Daddi-Moussa-Ider}\ \emph {et~al.}(2023)\citenamefont
  {Daddi-Moussa-Ider}, \citenamefont {Hosaka}, \citenamefont {Vilfan},\ and\
  \citenamefont {Golestanian}}]{PhysRevResearch.5.033030}%
  \BibitemOpen
  \bibfield  {author} {\bibinfo {author} {\bibfnamefont {A.}~\bibnamefont
  {Daddi-Moussa-Ider}}, \bibinfo {author} {\bibfnamefont {Y.}~\bibnamefont
  {Hosaka}}, \bibinfo {author} {\bibfnamefont {A.}~\bibnamefont {Vilfan}},\
  and\ \bibinfo {author} {\bibfnamefont {R.}~\bibnamefont {Golestanian}},\
  }\bibfield  {title} {\enquote {\bibinfo {title} {Axisymmetric monopole and
  dipole flow singularities in proximity of a stationary no-slip plate immersed
  in a brinkman fluid},}\ }\href@noop {} {\bibfield  {journal} {\bibinfo
  {journal} {Phys. Rev. Res.}\ }\textbf {\bibinfo {volume} {5}},\ \bibinfo
  {pages} {033030} (\bibinfo {year} {2023})}\BibitemShut {NoStop}%
\bibitem [{\citenamefont {Hosaka}, \citenamefont {Golestanian},\ and\
  \citenamefont {Daddi-Moussa-Ider}(2023)}]{hosaka2023hydrodynamics}%
  \BibitemOpen
  \bibfield  {author} {\bibinfo {author} {\bibfnamefont {Y.}~\bibnamefont
  {Hosaka}}, \bibinfo {author} {\bibfnamefont {R.}~\bibnamefont
  {Golestanian}},\ and\ \bibinfo {author} {\bibfnamefont {A.}~\bibnamefont
  {Daddi-Moussa-Ider}},\ }\bibfield  {title} {\enquote {\bibinfo {title}
  {Hydrodynamics of an odd active surfer in a chiral fluid},}\ }\href@noop {}
  {\bibfield  {journal} {\bibinfo  {journal} {arXiv preprint arXiv:2303.11836}\
  } (\bibinfo {year} {2023})}\BibitemShut {NoStop}%
\bibitem [{\citenamefont {Daddi-Moussa-Ider}\ \emph {et~al.}(2019)\citenamefont
  {Daddi-Moussa-Ider}, \citenamefont {Kurzthaler}, \citenamefont {Hoell},
  \citenamefont {Z{\"o}ttl}, \citenamefont {Mirzakhanloo}, \citenamefont
  {Alam}, \citenamefont {Menzel}, \citenamefont {L{\"o}wen},\ and\
  \citenamefont {Gekle}}]{daddi2019frequency}%
  \BibitemOpen
  \bibfield  {author} {\bibinfo {author} {\bibfnamefont {A.}~\bibnamefont
  {Daddi-Moussa-Ider}}, \bibinfo {author} {\bibfnamefont {C.}~\bibnamefont
  {Kurzthaler}}, \bibinfo {author} {\bibfnamefont {C.}~\bibnamefont {Hoell}},
  \bibinfo {author} {\bibfnamefont {A.}~\bibnamefont {Z{\"o}ttl}}, \bibinfo
  {author} {\bibfnamefont {M.}~\bibnamefont {Mirzakhanloo}}, \bibinfo {author}
  {\bibfnamefont {M.-R.}\ \bibnamefont {Alam}}, \bibinfo {author}
  {\bibfnamefont {A.~M.}\ \bibnamefont {Menzel}}, \bibinfo {author}
  {\bibfnamefont {H.}~\bibnamefont {L{\"o}wen}},\ and\ \bibinfo {author}
  {\bibfnamefont {S.}~\bibnamefont {Gekle}},\ }\bibfield  {title} {\enquote
  {\bibinfo {title} {{Frequency-dependent higher-order Stokes singularities
  near a planar elastic boundary: Implications for the hydrodynamics of an
  active microswimmer near an elastic interface}},}\ }\href@noop {} {\bibfield
  {journal} {\bibinfo  {journal} {Phys. Rev. E}\ }\textbf {\bibinfo {volume}
  {100}},\ \bibinfo {pages} {032610} (\bibinfo {year} {2019})}\BibitemShut
  {NoStop}%
\bibitem [{\citenamefont {Dalal}, \citenamefont {Farutin},\ and\ \citenamefont
  {Misbah}(2020)}]{Dalal2020}%
  \BibitemOpen
  \bibfield  {author} {\bibinfo {author} {\bibfnamefont {S.}~\bibnamefont
  {Dalal}}, \bibinfo {author} {\bibfnamefont {A.}~\bibnamefont {Farutin}},\
  and\ \bibinfo {author} {\bibfnamefont {C.}~\bibnamefont {Misbah}},\
  }\bibfield  {title} {\enquote {\bibinfo {title} {Amoeboid swimming in a
  compliant channel},}\ }\href@noop {} {\bibfield  {journal} {\bibinfo
  {journal} {Soft Matter}\ }\textbf {\bibinfo {volume} {16}},\ \bibinfo {pages}
  {1599--1613} (\bibinfo {year} {2020})}\BibitemShut {NoStop}%
\end{thebibliography}

\end{document}